\PassOptionsToPackage{usenames,dvipsnames,svgnames,x11names}{xcolor}
\documentclass[preprint,journal]{vgtc}                     
\vgtccategory{Research}

\vgtcpapertype{theory/model}

\RequirePackage[svgnames]{xcolor}

\usepackage{capt-of,etoolbox}
\usepackage{tcolorbox}

\usepackage{mathptmx}                  
\PassOptionsToPackage{warn}{textcomp}  
\usepackage{enumitem}
\usepackage{lettrine}
\usepackage{array}
\usepackage{soul} 
\usepackage{tikz}
\usepackage{pgfplots}
\pgfplotsset{compat=newest}
\usetikzlibrary{pgfplots.groupplots}
\usetikzlibrary{svg.path}
\usepgfplotslibrary{colorbrewer}
\usepackage{multirow}
\usepackage{adjustbox}
\usepackage{calc}
\usepackage[framemethod=tikz]{mdframed}
\usepackage{wrapfig}

\usepackage{pdfpages}

\usepackage{balance}
\usepackage[font={sf, footnotesize}]{caption}

\newrobustcmd{\phaseONEitem}{{\raisebox{\dimexpr-.6\height+.3\baselineskip}{{\includegraphics[width=3mm]{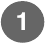}}}}}
\newrobustcmd{\phaseTWOitem}{{\raisebox{\dimexpr-.6\height+.3\baselineskip}{{\includegraphics[width=3mm]{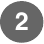}}}}}
\newrobustcmd{\phaseTHREEitem}{{\raisebox{\dimexpr-.6\height+.3\baselineskip}{{\includegraphics[width=3mm]{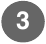}}}}}

\definecolor{myCorrectionColor}{RGB}{0,0,0}

\newrobustcmd{\TN}[1]{\textbf{T#1}}
\newrobustcmd{\TNN}[2]{\textbf{\color{black}}\textbf{#1}\color{black}}

\newrobustcmd{\GN}[1]{\textbf{G#1}}

\definecolor{mygreen}{RGB}{196,224,179}
\definecolor{myred}{RGB}{252,194,188}
\definecolor{myyellow}{RGB}{255,230,153}
\definecolor{mypurple}{RGB}{210,180,237} 
\definecolor{mygrey}{RGB}{218,218,218}
\definecolor{myblue}{RGB}{202,223,234}
\definecolor{skyblue}{rgb}{0.447,0.624,0.812} 
\definecolor{ABC}{RGB}{255,0,0}
\definecolor{JCR}{RGB}{0,0,205}
\definecolor{evfpurple}{RGB}{205,100,205}

\newcommand{\vignetteColour}{white}

\newlength{\myunit}
\newlength{\myunitTwo}
\newlength{\myraise}
\newcommand{\myTPerspective}[5]{%
  \begin{tcolorbox}[
    left=0mm,leftupper=0mm,leftlower=0mm,
    top=0mm,bottom=0mm,boxsep=1mm,
    middle=0mm,right=0mm,halign lower=left,
    sidebyside gap=5mm, lower separated=true,
    lefthand ratio=0.2,
    colback=\vignetteColour,colbacktitle=yellow!10!black,
     center title,
     before upper*=\fontsize{9}{9.8}\selectfont,
     before lower*=\fontsize{9}{10}\selectfont,
     after lower*=,
   ]   
\setlength{\intextsep}{3pt}
\setlength{\columnsep}{3pt}
\parindent0pt
\begin{wrapfigure}[#1]{L}{\myunit}
\vspace{#2}\includegraphics[width=\myunit]{#3}\vfill
\end{wrapfigure}
\color{black}\textbf{#4}~\color{black} #5%
 \end{tcolorbox}
}

\definecolor{darklavender}{rgb}{0.45, 0.31, 0.59}
\definecolor{darkviolet}{rgb}{0.58, 0.0, 0.83}

\newrobustcmd{\QN}[4]{\protect\noindent\textbf{\color{darkgray}\##1 #2 }(#3\color{Sepia}$\,\leftrightarrow$\,\color{black}#4)\color{black}.}

\newcommand{\myFsize}[6][1.2]{%
  \set@fontsize\baselinestretch{#2}{#2}%
  \set@fontsize{#1}\f@size\f@baselineskip%
  {\selectfont#3}
}
\definecolor{beige}{rgb}{0.96, 0.96, 0.86}
\definecolor{antiquewhite}{rgb}{0.98, 0.92, 0.84}
\definecolor{oldlace}{rgb}{0.99, 0.96, 0.9}

\newcommand*\numcircledtikz[1]{\tikz[baseline=(char.base)]{
            \node[shape=circle,draw,inner sep=1.2pt] (char) {#1};}}


\teaser{
  \centering
  \vspace{2mm}
  \includegraphics[width=\linewidth]{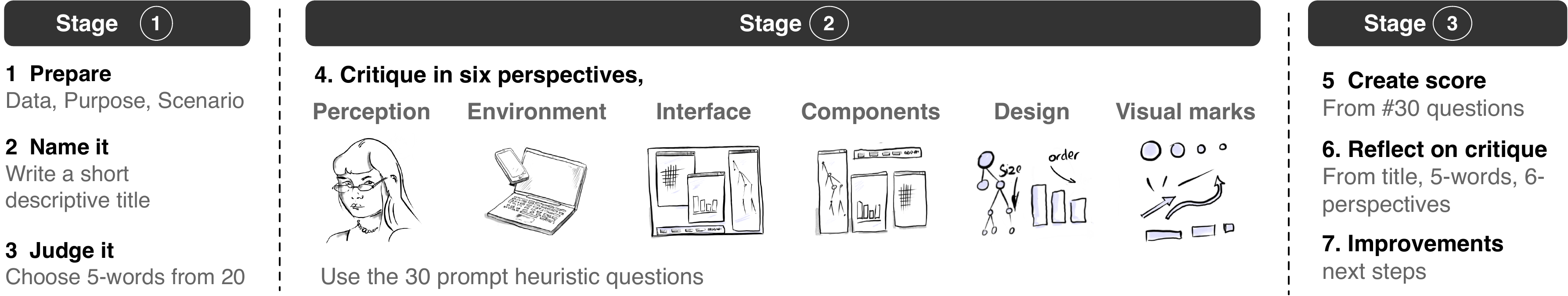}
  \caption{%
  	The Critical Design Strategy involves three stages: \texorpdfstring{\phaseONEitem}{1} overview, \texorpdfstring{\phaseTWOitem}{1} detail, and \texorpdfstring{\phaseTHREEitem}{1} review. The  appraiser---often the creator, developer, or designer---first considers the design holistically, summarising it and selecting five keywords from a set of twenty. They then evaluate it in depth across six perspectives, using 30 heuristic questions or semantic differential word pairs (opposite adjectives). Finally, the appraiser reflects on the critique, assigns an indicative score, and identifies areas for improvement.
  }  \vspace{2mm}
  \label{fig:PartsOfThePaperFIG}
}

\title{Critical Design Strategy: a Method for Heuristically Evaluating Visualisation Designs}%

\author{%
  \authororcid{Jonathan~C. Roberts}{0000-0001-7718-3181}, 
  \authororcid{Hanan Alnjar}{0000-0002-5123-4638},
  \authororcid{Aron E.\ Owen}{0000-0001-5660-5867} and
  \authororcid{Panagiotis~D. Ritsos}{0000-0001-9308-3885}
}

\authorfooter{
  \item
  	Jonathan~C. Roberts, Aron E. Owen and Panagiotis~D. Ritsos are with Bangor University.
  	E-mail: \{j.c.roberts\,\ aron.e.owen\,\ p.ritsos\}@bangor.ac.uk.
  \item
  	Hanan Alnjar is with University of Basrah, and previously Bangor University, h.r.alnjar@uobasrah.edu.iq
}
 
\abstract{
We present the Critical Design Strategy (CDS)---a structured method designed to facilitate the examination of visualisation designs through reflection and critical thought. The CDS helps designers think critically and make informed improvements using heuristic evaluation. When developing a visual tool or pioneering a novel visualisation approach, identifying areas for enhancement can be challenging. Critical thinking is particularly crucial for visualisation designers and tool developers, especially those new to the field, such as studying visualisation in higher education. The CDS consists of three stages across six perspectives: Stage 1 captures the essence of the idea by assigning an indicative title and selecting five adjectives (from twenty options) to form initial impressions of the design. Stage 2 involves an in-depth critique using 30 heuristic questions spanning six key perspectives---user, environment, interface, components, design, and visual marks. Stage 3 focuses on synthesising insights, reflecting on design decisions, and determining the next steps forward. We introduce the CDS and explore its use across three visualisation modules in both undergraduate and postgraduate courses. Our longstanding experience with the CDS has allowed us to refine and develop it over time: from its initial creation through workshops in 2017/18 to improvements in wording and the development of two applications by 2020, followed by the expansion of support notes and refinement of heuristics through 2023; while using it in our teaching each year. This sustained use allows us to reflect on its practical application and offer guidance on how others can incorporate it into their own work.
}

\begin{document}
\renewcommand*{\backref}[1]{
  %
}
\maketitle
\section{Introduction}
\newcommand{\myLikert}{$\circ\circ\circ\circ\circ$}

One of the most important skills that a visualisation designer needs to possess is that of critical thinking. Creating effective visualisations is a combination of technical skills, domain knowledge, and attention to detail, along with design and critical thinking. There are many ways to improve technical skills. Proficiency in data analysis and use of visualisation software such as Tableau, ggplot, D3.js or matplotlib can be gained through tutorials, workshops, classes, and books, and so forth. Domain knowledge can be learnt, or experts involved may provide it. Yet, there are few methods that help people think critically over their visualisation designs. Critical thinking methods outlined in the literature typically consist of broad catalogues of overarching concepts. Additionally, due to the ad hoc nature of design, it is often difficult for people to know how to structure their critical thought process. Indeed, people must consider different designs, layouts, and visualisation arrangements, while gauging the suitability of each option in the use-case they set to tackle. 

The requirement for having such skills is becoming more crucial to a wider community of people, as data becomes more pervasive across various industries and sectors.  The demand for individuals with strong critical thinking and visualisation design skills is growing rapidly. 
In particular, students of various levels and disciplines, seek to learn how to create data visualisations, as means of analysis and support for their work.  While expert visualisation developers rely on their knowledge, and experience of what works (or does not), learners struggle to understand where to start, or how to organise their critical thinking. Heuristic guidelines~\cite{Nielsen:1990:HEU:97243.97281,zuk2006theoretical,forsell2012evaluation}, where people evaluate a tool against a set of recognised usability principles (the \textit{heuristics}) can help. But even with these strategies, learners can find it difficult to critique designs, due to lack of experience, and often struggle to know how to go about critiquing designs in an systematic way.

To address these challenges, we present the Critical Design Strategy (CDS), extending our IEEE VIS 2023 poster presentation~\cite{RobertsETAL2023CDSPoster}. 
Consisting of three stages, and several thought-provoking segments, the CDS aids individuals in critically reflecting on visualisation designs. It serves as a valuable tool to structure critical thinking for design visualisation. It is especially useful for educators in the classroom, facilitating the generation of insights that can be leveraged to craft critique reports (documents that provide feedback on a specific design, which offer constructive criticism and recommendations for improvement). Stage~\texorpdfstring{\phaseONEitem}{1} of the strategy gets individuals to consider the design holistically. In Stage \texorpdfstring{\phaseTWOitem}{2} they dive into detail, and systematically consider 30 aspects of the design (from six perspectives). In Stage \texorpdfstring{\phaseTHREEitem}{3} individuals calculate an average score, reflect on their critique and decide what to do next.

We contribute:
(i) a broad discussion on the rhetoric of critique with a particular focus on visual critique  (\cref{SEC:RelatedWork}), 
(ii) a detailed breakdown of the Critical Design Strategy (\cref{SEC:TheCDS}), including an in-depth exploration of each of the three stages, six perspectives and 30 heuristic questions. In addition, we provide supplementary material with comprehensive notes that can be used by teachers.
(iii) The design and evolution of the CDS, from its inception in 2017/18 through multiple workshops to its adaptations over the years (\cref{SEC:DesignOfCDS}, and \cref{FIG:evolution}). This section includes several evaluations, including an assessment of the use of the 20-words, which led to refinements, an initial evaluation with 30 students, and a review of the latest version.
(iv) Finally, we reflect on its use in our teaching across eight cohorts (see \cref{FIG:evolution}) and highlight key focus areas to help others adopt and apply it in their own work (\cref{SEC:DiscussionConclusions}).

\section{Background}
\label{SEC:Background}
We are strong advocates for using (and developing) \textit{explanatory frameworks}. In the context of science education, explanatory frameworks are particularly relevant for helping learners develop a deep understanding of scientific concepts~\cite{TreagustHarrison2000_ExplanatoryFrameworks}. They can be found across domains, spanning cell theory, evolutionary theory, energy transfer and plate tectonics. 
The visualisation community also emphasises their research, e.g., Bach et al~\cite{BachETALChallenges2024} call for theoretical frameworks tailored to visualisation education and those offering practical guidelines.
Explanatory frameworks serve as guiding structures for individuals to navigate and comprehend diverse subjects effectively, and as roadmaps that help learners organise thoughts, explore information systematically, and understand complex concepts more easily. 
In visualisation there are several explanatory frameworks, such as Bertin's method of mapping data onto visual properties~\cite{Bertin1983}, the data-flow paradigm, Shneiderman's visual information-seeking mantra~\cite{shneiderman1996eyes}, Amar and Stasko's knowledge and task-based framework~\cite{AmarStasko2004}, and Munzner's nested model~\cite{Munzner2009}. 
This work builds on three of our previous \textit{explanatory frameworks}; our Explanatory Visualisation Framework (EVF)~\cite{RobertsETAL2018_EVF}, a design and build strategy for courses where people create explanatory visualisations, the Five Design-Sheets (\textbf{FdS})~\cite{RobertsHeadleandRitsos15_TVCG,RobertsHeadleandRitsos2017_BOOK}, a method to consider alternative designs through sketching, and the Critical Thinking Sheet (CTS)~\cite{RobertsRitsos2020CTS}, a strategy to encourage learners to critically think, and sketch their algorithm, before coding.

We started to develop the Critical Design Strategy (CDS) around 2015, while reflecting on critical thinking, design and pedagogical approaches in general. Our pedagogic strategy is two-fold. First, learners choose a dataset, analyse it, and perform a low-fidelity design study -- we use the FdS, which they write as a design report. Second, learners develop a prototype visualisation based on their design and submit both their implementation and a reflective report, and in most cases we request students to use \href{http://processing.org}{Processing.org} to create their visualisations. We chose Processing to encourage the development of unique and innovative visual solutions rather than relying on predefined visualisation types such as bar charts or line graphs.
Enrolment has increased from around 30 students in 2017-2020 to 70 in 2025. Classes typically combine lectures with practical tasks that reinforce learning and allow us to observe student understanding. We found that many learners struggled to critically evaluate visualisations and organise their critique reports. For example, in one group exercise we task students to find and critique visualisation examples online. Their feedback often reflected personal preference, such as ``I like it'' or ``it looks good'', rather than applying objective principles taught in previous lectures, including design thinking~\cite{ware2012information}, task analysis~\cite{shneiderman1996eyes}, retinal variables~\cite{Bertin1983}, Gestalt theory~\cite{Munzner2014Book,RobertsHeadleandRitsos2017_BOOK}, and established design guidelines~\cite{DeiterRams}. 
Furthermore when grading their work, we consistently provided feedback, urging learners to \textit{``justify your decisions''} and ``\textit{structure your critical thinking report}'' as we often found their reports lacked clear organisation.
With critical thinking, learners need to (1) ask questions, (2) collect information, (3) contemplate alternative potential solutions, (4) understand and empathise different viewpoints, and finally (5) communicate in a clear way~\cite{Facione1990critical}. Learners were not doing any of these tasks effectively. What is needed is a formal method and a set of steps to help people critique their ideas in a systematic way. 
The CDS provides an `explanatory framework'~\cite{TreagustHarrison2000_ExplanatoryFrameworks}, with three core parts (overview, detail, review), see~\cref{fig:PartsOfThePaperFIG}, and a structure for results reporting and discussion.

We introduced the CDS in our teaching around 2017~\cite{hanan2017}, refined the questions and structure until 2020, and published a IEEE VIS 2023 poster presentation~\cite{RobertsETAL2023CDSPoster}.  We give two lectures: one on critical thinking, and another specifically on the CDS. 
We have used this strategy with an Information Visualisation (InfoVis) module; a 20 credit compulsory module for an MSc in Data Science, and optional for MSc in Advanced Computer Science (CS) and BSc Computer Science (MSc version is 180 credits long). We estimate over 300 students have used the CDS. In our classes, students use the CDS (at least) twice. First, after ideating designs using the FdS, they perform a CDS critique on their realisation design (FdS sheet~5). Second, they critique their implementation using the CDS, and write a reflective report following the CDS structure (see \cref{SEC:TheCDS}, \cref{FIG:evolution}).

\section{Related Work}
\label{SEC:RelatedWork}
Critical analysis is at the heart of academic tradition of reason and argument and is taught across the curriculum~\cite{PaulElder2009,Biggs2011teaching}. We begin with rhetoric and critique, then move to visualisation. 

\subsection{Rhetoric and critique}
\label{SSEC:Rhetoric}

Especially for writing critical essays, the art of contrastive rhetoric writing, which inspires this work, encourages the reviewer to break the work into individual parts, and identify strategies that the author has used to persuade the audience.
Likewise, in a design context, we are breaking the visual depiction into different categories, considered individually, to make an overarching judgement on the design. 
Influenced by Aristotle, Cicero described five canons of rhetoric: \textit{inventio} (invention/creation), \textit{dispostio} (arrangement), \textit{elocutio} (style), \textit{memoria} (memory/recollection) and \textit{pronuntiatio} (delivery)~\cite{selzer2004rhetorical}, which, although intended for public speaking, describe the writing process as well.
These terms can also be mapped to the Aristotle's means of persuasion, of \textit{ethos}, \textit{pathos} and \textit{logos}: detailing the trustworthiness and credibility of the orator, how they use emotion to tender support and their logical argument. These subsections influence the six perspectives of the CDS. 
Consequently, there are several skills that learners aspiring to be critical thinkers require~\cite{Halpern1999teaching,Ennis1962concept,ennis2011critical,OREILLY2022101110,calma2021assessing}, such as being informed, honest and open, orderly, and not shying from hard work. Facione writes: ``\textit{The ideal critical thinker is habitually inquisitive, well-informed, .. 
honest in facing personal biases, prudent in making judgements, willing to reconsider, clear about issues, orderly in complex matters, .. 
	focused in inquiry, and persistent..
    }''~\cite{Facione1990critical}.

While we now have a written, rather than oral, rhetoric tradition, these ideas have been refined and taught across levels in our education system, for all types of learning, and different critical thinking tasks. 
Acronyms such as SOAPSTone are useful, for writers to consider the Speaker, Occasion, Audience, Purpose, Subject and Tone of a written document. 
There are several well-known educational models that include critical reflection, such as: a) Borton's~\cite{borton1970} `what', `so what', and `now what', b) the five W's \textit{who}, \textit{what}, \textit{why}, \textit{when}, \textit{where}~\cite{eppler2007visual}, and c) Roberts' et al. should-you, could-you, what-if-you~\cite{RobertsHeadleandRitsos2017_BOOK}. Many of these models stem from problem solving (e.g., Polya~\cite{Polya2014solve} and Duncker~\cite{Duncker1945problem}), whereas
educational concepts are often influenced by Dewey~\cite{DeweyJohn_1910}, 
David Kolb's reflective model of experiential learning~\cite{kolb2014experiential}, and Ennis'~\cite{ennis2011critical} on dimensions of critical thinking (\textit{logical}, \textit{criteria} and \textit{pragmatic}). More importantly, critical thinking should not only happen at the end of the design process. As Borton~\cite{borton1970} suggests, it needs to take place when considering past activities, while something is taking shape and for anticipating what could happen in the future. While useful, however, these structures are not design or visualisation specific.

\subsection{Visual critique and visualisation}
Design reflection is an important aspect of general interface design and therefore visualisation tool design. There are many ways to evaluate interfaces, including 
\textit{automatically} (using algorithms and metrics), \textit{empirically} (assessed by real users~\cite{CHEN:2000:ESI:362757.362818}),
\textit{formally} (using formulas and usability measures), and
\textit{informally} (based on the skill and experience of evaluators~\cite{Nielsen:1990:HEU:97243.97281}). Alternative evaluation methods each have trade-offs. Expert reviews require access to skilled evaluators and can lack consistency. Heuristic evaluations are common but often informal and difficult for learners to apply without guidance. Empirical user studies yield valuable insights but require time, participants, and ethics approval. Automated metrics offer speed but may overlook contextual or qualitative aspects. In contrast, the CDS provides a structured and accessible approach suited to formative evaluation during design. It falls under the informal classification; a heuristic method, where people evaluate the output against predefined heuristics~\cite{Nielsen:1990:HEU:97243.97281}. 
We suggest the CDS process be used formatively alongside the design process, for example: ideate with the FdS~\cite{RobertsHeadleandRitsos15_TVCG}, perform a CDS on the proposed design, adjust the design, implement, and re-evaluate with CDS. 

Kosara~\cite{Kosara2007} explains that critical inspection is a useful form of evaluating visualisations and writes ``\textit{visualisation criticism could be a tool for further developing and increasing the usefulness of visualisation theory}''~\cite{KosaraEtAl2008} but does not offer a formal structure. We already have strategies that help us consider alternative visualisation designs, such as the FdS method~\cite{RobertsHeadleandRitsos15_TVCG}, sketching user interfaces (e.g., Buxton~\cite{buxton2010sketching}), and Munzner's Nested Model~\cite{Munzner2009}, which provide guidance how to develop and consider visualisations. To facilitate \textit{creative thinking} for visualisation design, we need to employ \textit{critical thinking}. Critical thinkers seek to find alternative perspectives that they analyse in an open-minded way, avoiding hasty decisions~\cite{PaulElder2009}. If we are unable to assess the products we create and produce, and ascertain what to change, then we will not be able to improve our solutions.
Yet, getting towards this goal is not only about considering alternatives (divergent thought) but also employing both convergent and critical thinking approaches to reach a suitable solution~\cite{Polya2014solve}. Inevitably, critical thinking must take place throughout the whole design process, from deciding what data to select, how to process and enhance the data, to what visual design to choose, etc. Thus, developers need to simultaneously employ bottom-up and top-down thinking; be open to alternative potential solutions (however unusual they are), while concurrently envisioning the end output, how to create it, and judge whether the problem at hand is solved~\cite{Duncker1945problem}.

Bottom-up methods~\cite{RobertsHeadleandRitsos15_TVCG, buxton2010sketching}) may help people 
ideate alternatives, but the process is still challenging. It is difficult to be open-minded, unbiased, put personal opinions aside, and it is too easy fixate on one solution (c.f., Facione~\cite{Facione1990critical}). It can be problematic to know how to organise the information appropriately to make the right decisions. 
Working together in a group can help to moderate views. Jackson et al.\ \cite{Jackson:2012:TMM:2442576.2442580} employ a group critiquing process that is integrated with sketching and design. We, likewise, integrate critical analysis with the design, but in our case the individual critic (rather than a group) makes judgements on the visualisation, using our formal structure for the critical process. Saraiya et al.\ \cite{SaraiyaEtAlInsight2005} also use insight-based methods, and provide a structure of eight characteristics: observation, time, domain value, hypothesis, directed vs. unexpected insights, correctness, breadth vs. depth and categorisation. This latter work is the closest to ours, yet we develop a more systematic structure, define a set of question that can be readily taught, learnt, and used in various situations including critique assignments. The CDS also delivers a result (score), which can help people to compare compare previous critical analysis of visualisations, and indicate improvement. 

Top-down critical thinking also requires people to imagine how a potential solution would appear, how it would be utilised in the intended environment, and how it could work using available technologies. Experts, when characterising the problem~\cite{Munzner2009}, build up a broad picture of solutions in their mind. They draw on past experiences, are able to abstract ideas and create many visions of potential futures, to decide on the right result~\cite{SwellerEffectsOnLearning}. But for learners, this is much more challenging~\cite{Halpern1999teaching}. They do not have the structures, schema and cannot rely on years of experience. Learners need \textit{explanatory frameworks}~\cite{TreagustHarrison2000_ExplanatoryFrameworks} to facilitate  the systematic exploration and interpretation of the idea. Thus, as researchers, we need to create structures and methods to support people in their critique.

Informal styles of evaluation are useful, especially to help focus and organise personal thoughts during the design process and to aid critical thinking~\cite{BrathBanissi2016}. Amar and Stasko~\cite{AmarStasko2004} walk the user through thinking about rationales and tasks, and Munzner~\cite{Munzner2009} implicitly advocates critical thinking throughout the whole visualisation creation process. In fact, ideas of visual inspection have a long history. Bertin~\cite{Bertin1983} expounds three levels to read a graphic, from the elementary (looking at visual variables and marks), to look at patterns within the presentation, to observing the whole. Cleveland~\cite{cleveland1993model} also describes three perception operations: detection, assemble and estimation. Lohse~\cite{Lohse1990classifying} uses similar ideas to help classify visual representations, and develops a model to understand graphical depictions~\cite{Lohse:1993:CMU:1461776.1461779}, while inspection techniques were used by Conversy et al.\ \cite{ConversyChattyHurter2011} to observe visualisations. Workshops can also help people discuss opportunities and challenges with visualisations~\cite{KerznerETAL2019}. 
Additionally, visualisation design guidelines are detailed in many books (e.g., \cite{Ware12,Munzner2014Book,RobertsHeadleandRitsos2017_BOOK}), yet while these books provide a rich tapestry of information, it is difficult for learners to aggregate this information into a single critical review structure. 

There is a growing interest in evaluating visualisations~\cite{chen2000empirical,plaisant2004challenge,Andrews:2006:EIV:1168149.1168151,Sedlmair:2012:DSM:2720013.2720398,Lam:2012:ESI:2360755.2361044,InsenbergEtAl2013, Wall_2019}. However, most current techniques require that the tool be fully built and evaluate the usability or user experience of the tool~\cite{Lam:2012:ESI:2360755.2361044}. Several researchers have produced classifications on the different types of evaluation strategies. For instance, Plaisant~\cite{plaisant2004challenge} focused on four areas: (i) controlled experiments, (ii) tool usability, (iii) controlled experiments comparing tools and (iv) case studies of tools in settings. 
Isenberg et al.\ \cite{InsenbergEtAl2013} extended the seven scenarios of Lam et al.\ \cite{Lam:2012:ESI:2360755.2361044} offering a comprehensive review of visualisation evaluation methods. Based on their framework, the CDS aligns with Understanding Environments and Work Practices (UWP), as it prompts users to consider how a tool fits within its intended context. When used during iterative refinement, the CDS also supports Evaluating Visual Data Analysis and Reasoning (VDAR) as an informal  approach. 

There are several works on heuristics in visualisation, highlighting benefits in designer/expert collaboration~\cite{shneiderman2006strategies,ToryMoller2005}.
Heuristics can be general, e.g., ``\textit{Consider people with colour blindness}'' or specific ``\textit{ensure visual variable has sufficient length}'', or ``\textit{provide multiple levels of detail}''~\cite{zuk2006theoretical}. Many researchers focus on ten (like Nielsen~\cite{Nielsen:1990:HEU:97243.97281}) while others aggregate several lists. For example,  Forsell and Johnson present ten~\cite{ForsellJohansson2010,forsell2012evaluation}, 
Zuk and Carpendale present twelve~\cite{zuk2006theoretical} drawn from Tufte~\cite{tufte1983visual}, Ware~\cite{Ware12} and Bertin~\cite{Bertin1983}.
Engelbrecht et al.~\cite{engelbrecht2014information} aggregates a list.
Scholtz~\cite{scholtz2011developing} explains that evaluators of the VAST challenge drew heuristics from Zuk and Carpendale~\cite{zuk2006theoretical}, Shneiderman's visual information seeking mantra~\cite{shneiderman1996eyes} and Amar and Stasko's knowledge and task-based framework~\cite{AmarStasko2004}. Other heuristic lists include~\cite{dowding2018development,vaataja2016information,de2017adapting,tarrell2014toward}).
Heuristics help people think, but they are typically not encapsulated in a formal structure. One exception is Wall et al.~\cite{Wall_2019} who structure them around \textit{insight}, \textit{time}, \textit{essence} and \textit{confidence}~\cite{stasko2014value}. Additionally, Eppler and Burkhard~\cite{eppler2007visual} organises them by \textit{what}, \textit{why}, \textit{who}, and \textit{when}. Our approach is also structured, but we follow a double-diamond approach~\cite{council2007eleven} of 
starting broad (summary), delving deep (30 questions in six perspectives), and a final general step (reflection).

\section{The Critical Design Strategy (CDS)}
\label{SEC:TheCDS}
The final incarnation of the CDS comprises of three stages, carried out sequentially: \textbf{overview}, \textbf{detail} and \textbf{review},~\cref{fig:PartsOfThePaperFIG}.  In this section, we describe the purpose of each stage, outline the overall process, and use vignettes to summarise key steps and highlight important points.
We begin by defining the key roles. The \textit{appraiser} is the individual (or group) conducting the critique; typically the creator, developer, or designer of the work. 
The subject of appraisal is a visualisation \textit{artefact}, which may take the form of a sketch, an interactive tool, a physicalisation, or even a poster. This artefact presents \textit{data}, is crafted by a \textit{designer}, coded by a \textit{developer}, and ultimately intended for use by a \textit{user}.
These roles may be carried out by different individuals or fulfilled by the same person; for example, a learner might design a data visualisation, code it, and appraise it with the CDS.

\vspace{1mm}
\begin{itemize}[nosep,leftmargin=\parindent, align=left, labelwidth=\parindent, labelsep=2pt]
    \item[\texorpdfstring{\phaseONEitem}{1}] {\textbf{Overview}}. After suitable preparation, 
    assign a name, summarise its essence, and holistically critique by selecting five words.  
    \item[\texorpdfstring{\phaseTWOitem}{2}] {\textbf{Detail}}.
   Critique artefact. The appraiser conducts a thorough critique by responding to 30 heuristic questions across six perspectives.
    \item[\texorpdfstring{\phaseTHREEitem}{3}] {\textbf{Review}}. Finally, the appraiser reflects on both the overall critique and the detailed analysis to identify the next steps. 
\end{itemize}

\subsection{Stage~\texorpdfstring{\phaseONEitem}{1} -- Overview}
The primary objective of the first stage is to ensure a thorough understanding of the topic and to make holistic assessments of the artefact. Critical thinking necessitates individuals to be ``\textit{well-informed}''~\cite{Facione1990critical}. Individuals should adequately \textbf{prepare} and ensure a thorough understanding of both the challenge and associated data. Data visualisation cannot be pursued without access to data. It crucial to consider the composition of the data and its organisational aspects, such as sparsity and structure. This involves identifying variables, understanding their nature (categorical, ordinal, quantitative, etc.), and recognising the purpose for which the data was collected. Additionally, comprehending the main objective of the visualisation and the intended user tasks is essential. Contextual information, including the creator's intent and the environment in which the visualisation will be utilised, should be understood to ensure effective use. 

To confirm understanding, the appraiser \textbf{name} the artefact/design, and summarise its \textbf{essence}. 
The act of naming the design commences the critical thinking process. Crafting a brief, concise title (of two or three words) compels consideration of what is crucial~\cite{RobertsHeadleandRitsos15_TVCG}. 
The holistic critique continues, by \textbf{circling five} of the twenty words (\cref{tab:20-first-impression-words-final}). This task acts as a preliminary, intuitive assessment. While such initial assessments may be wrong, they are reviewed in the Review stage~\texorpdfstring{\phaseTHREEitem}{3}.

\begin{figure}[t]            
  \begin{tcolorbox}[title,
    left=0mm,leftupper=0mm,leftlower=0mm,
    top=0mm,bottom=0mm,boxsep=1mm,
    middle=0mm,right=0mm,halign lower=left,
    sidebyside gap=5mm, lower separated=true,
    lefthand ratio=0.2,
     colback=\vignetteColour,colbacktitle=yellow!10!black,
     title={Stage~\numcircledtikz{1} -- Overview},
     center title,
     before upper*=\fontsize{8}{9.8}\selectfont,
     before lower*=\fontsize{8}{9}\selectfont,
     after lower*=,
   ]
      
\textbf{Assign a name to the design}: 
   
   \textbf{Summarise essence:}
   \tcblower
     \textbf{Circle 5 (first impression) words:}\vspace{2mm}
     
     \raggedright clear confusing sensible indifferent clever reliable pointless indistinctive complex organised moderate spectacular useless average bad fulfilling useful fair vague beautiful 
     \end{tcolorbox}
\centering
\vspace{-2mm}
        \caption{Following adequate preparation, assign name, summarise essence, conduct a holistic critique by selecting five descriptive words.}\vspace{-3mm}
        \label{tab:20-first-impression-words-final}

\end{figure}

\definecolor{lightgray}{gray}{0.9}
\begin{figure*}[h]
\begin{tcolorbox}[title,
    left=0mm,leftupper=0mm,leftlower=0mm,
    top=0mm,bottom=0mm,boxsep=1mm,
    middle=0mm,right=0mm,halign lower=left,
    sidebyside gap=5mm, lower separated=true,
    lefthand ratio=0.2,
     colback=\vignetteColour,colbacktitle=yellow!10!black,
     title={Stage~\numcircledtikz{2} -- Detail} (comprehensive critical evaluation),
     center title,
     before upper*=\fontsize{9}{9.8}\selectfont,
     before lower*=\fontsize{9}{10}\selectfont,
     after lower*=,
   ]

\fontsize{7}{8.2}\selectfont
    \begin{tabular}{l@{\hspace{2mm}}l@{\hspace{2mm}}lc|lll}
    \textbf{Perspective}&&\textbf{Heuristic question}&-2 -1 0 1 2&\multicolumn{3}{l}{\textbf{Range of answers, from poor to good}}\\
    \hline&&&&\\[-.4em]
\textbf{User}&~~\#1&Is suitable for the user and task&\myLikert&Unsuitable&$\leftrightarrow$&Suitable\\
\multirow{4}{*}{\includegraphics[width=\myunitTwo]{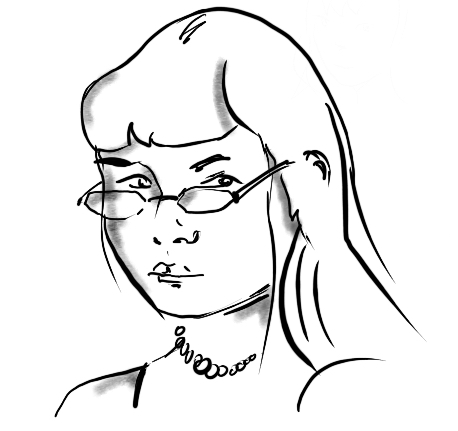}}&~~\#2&Is understandable for user and task to hand&\myLikert&Incomprehensible&$\leftrightarrow$&Understandable\\
&~~\#3&It doesn't require guesswork&\myLikert&Requires guesswork&$\leftrightarrow$&Clear assumptions\\
&~~\#4&Is trustworthy&\myLikert&Distrustful&$\leftrightarrow$&Trustful\\
&~~\#5&Would be useful&\myLikert&Useless&$\leftrightarrow$&Useful\\
   & & & \\[-0.4em]
\textbf{Environment}&~~\#6&It would fit in with other technologies&\myLikert&Wrong setting&$\leftrightarrow$&Right setting\\
\multirow{4}{*}{\includegraphics[width=\myunitTwo]{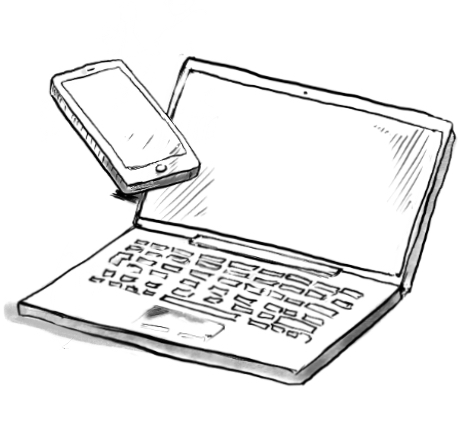}}&~~\#7&Uses suitable technology&\myLikert&Unsuitable technology&$\leftrightarrow$&Right technology\\
&~~\#8&Has appropriate interaction&\myLikert&Unsuitable interaction&$\leftrightarrow$&Appropriate interaction\\
&~~\#9&Its sizing is correct &\myLikert&Unsuitable size&$\leftrightarrow$&Suitable physical size\\
&\#10&Gives a positive ambience&\myLikert&Poor vibe/ambience&$\leftrightarrow$&Positive ambience\\
   & & & & \\[-0.4em]
\textbf{Interface}&\#11&Suitable user interface&\myLikert&Unsuitable GUI&$\leftrightarrow$&Suitable GUI\\
\multirow{4}{*}{\includegraphics[width=\myunitTwo]{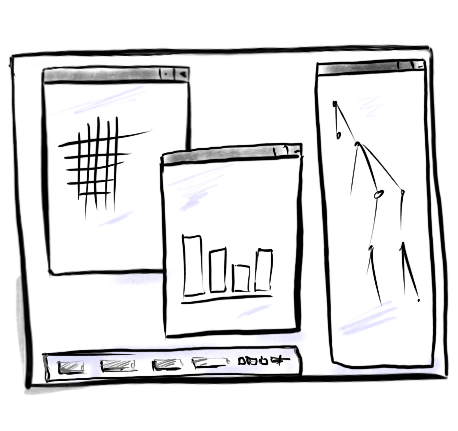}}&\#12&Ergonomic interface&\myLikert&Uncomfortable&$\leftrightarrow$&Ergonomic\\
&\#13&Facets are sized suitably&\myLikert&Poorly proportioned&$\leftrightarrow$&Suitable sized facets\\
&\#14&Interface suitably spaced &\myLikert&Poor facet spacing&$\leftrightarrow$&Relevant spacing\\
&\#15&Suitable quantity of interface parts&\myLikert&Unsuitable facet quantity&$\leftrightarrow$&Suitable facet quantity\\
   & & & & \\[-0.4em]
\textbf{Components}&\#16&Has all necessary components&\myLikert&Missing components&$\leftrightarrow$&All necessary components\\
\multirow{4}{*}{\includegraphics[width=\myunitTwo]{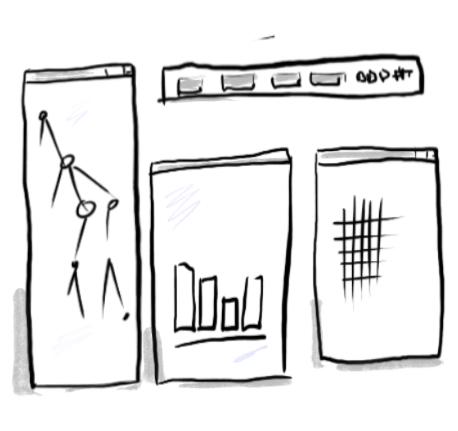}}&\#17&Has all suitable output/view types&\myLikert&Unsuitable types&$\leftrightarrow$&Suitable view types\\
&\#18&Clear relationships between parts&\myLikert&Unclear correspondences&$\leftrightarrow$&Clear view relationships\\
&\#19&Task can be easily performed&\myLikert&Task unfulfilled&$\leftrightarrow$&Task easily performed\\
&\#20&Suitable organisation of components&\myLikert&Poor component layout&$\leftrightarrow$&Good component layout\\
   & & & & \\[-0.4em]
\textbf{Design}&\#21&Inspiring design&\myLikert&Uninspiring&$\leftrightarrow$&Inspiring\\
\multirow{4}{*}{\includegraphics[width=\myunitTwo]{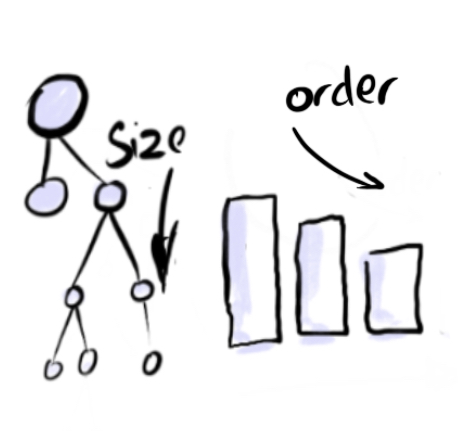}}&\#22&Aesthetic and visually attractive&\myLikert&Unattractive&$\leftrightarrow$&Visually attractive (aesthetic)\\
&\#23&Good composition and space utilisation&\myLikert&Poor layout&$\leftrightarrow$&Good composition\\
&\#24&Suitable coverage of data/underpinning facets/concepts&\myLikert&Unsuitable coverage&$\leftrightarrow$&Suitable coverage\\
&\#25&Clear instructions, labels, legends to give context&\myLikert&Poor labels/legends&$\leftrightarrow$&Suitable legends/labels\\
   & & & & \\[-0.4em]
\textbf{Visual marks}&\#26&Right choice of channels to communicate things clearly&\myLikert&Poor choice of channels&$\leftrightarrow$&Good channel choices\\
\multirow{4}{*}{\includegraphics[width=\myunitTwo]{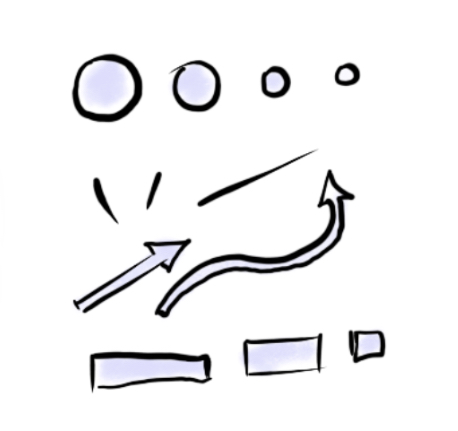}}&\#27&Communicates appropriate relationships/morphisms&\myLikert&Inappropriate mappings&$\leftrightarrow$&Appropriate mappings\\
&\#28&The types of marks used, communicate things well&\myLikert&Inappropriate mark types&$\leftrightarrow$&Suitable mark types\\
&\#29&Components are shown at the right level of abstraction/detail&\myLikert&Poor scale/zoom&$\leftrightarrow$&Good scale/zoom\\
&\#30&Nothing is hidden that shouldn't be hidden&\myLikert&Overplotting&$\leftrightarrow$&Clear display, easy read\\
    \end{tabular} 
\end{tcolorbox}
    \centering
        \vspace{-2mm}\caption{Conduct a comprehensive critical evaluation of the artefact/design. Follow the questions (in the six perspectives: User, Environment, interface, components, design, marks), recording the answers in the Likert scale. Make notes that justify your decisions.  
        }
        \vspace{-2mm}
    \label{fig:stage2-detail}
\end{figure*}

\smallskip
\subsection{Stage~\texorpdfstring{\phaseTWOitem}{2} -- Detail}

During the second stage, the aim is for the appraiser to conduct a comprehensive critique (\cref{fig:stage2-detail}), by considering 30 heuristic questions in six perspectives (\textbf{User, Environment, Interface, Components, Design, Visual Marks}), each with its own Likert scale. These six perspectives encourage a top-down approach in the critique, with the questions themselves encouraging reflection while providing a structure for appraising different design viewpoints. The critique of this stage can be written as a report (as when we use the CDS in assessments). An overall score can be calculated at this stage. We detail some of the possible questions to reflect upon at this stage in our Supplementary material.\\[-0.7em]

\newcommand{\myPerspectiveIII}[1]{%
\begin{wrapfigure}[4]{l}{1.2cm}
    \vspace{-6mm}
\begin{tcolorbox}[width=1.3cm,
    left=0mm,leftupper=0mm,leftlower=0mm,
    top=0mm,bottom=0mm,boxsep=1mm,
    colback=\vignetteColour,colbacktitle=yellow!10!black,
   ]       
      \includegraphics[width=1.0cm]{#1}\end{tcolorbox}
\end{wrapfigure} 
}
\myPerspectiveIII{pictures/pic1-}
\noindent \textbf{User:} The first step is to consider the user's perspective and empathise with their point of view, and expectations in terms of their skills and experience. Topics to reflect upon include task suitability, how understandable the visualisation is, its trustworthiness and usefulness, echoing more traditional sub-dimensions of User Experience.

\myPerspectiveIII{pictures/pic2-}
\noindent \textbf{Environment:} The next step is to consider the suitability of the proposed environment for end usage. This needs to consider the overall scenario of usage, where will this take place physically, access to different technologies and their potential interoperability. 
The appraiser should consider whether the design of the artefact is appropriate and well-suited for the intended purpose, environment, need, and with appropriate ergonomics? E.g., a static display for an e-book, while a 3D environment is suitable for an immersive head-mounted display.
Likewise, considerations should assess what interaction affordances are available for sensemaking in terms of the environment. The term \textit{interaction} should be understood in the appropriate context. Rather than simply evaluating whether it is an interactive Human-Computer Interaction (HCI) tool (e.g., using a mouse), assess whether the level and mode of interaction align with the specific situation and usage context. For example, if the artefact is a visualisation poster, interaction occurs when people physically move closer or further away from it.
Ergonomics questions should be considered, such as whether physical interaction can take place, say by moving closer/away from a powerwall~\cite{AndrewsNorth2013}, but environmental obstacles may impede such interaction.

\myPerspectiveIII{pictures/pic3-}
\noindent \textbf{User Interface:} Following the critique on the usage environment, the appraiser should reflect on the user interface available. This is more often an interactive, programmed interface encompassing different interactive visualisation techniques, often employing specific hardware (e.g., HMDs). However, the interface can also be physicalisations, printed outputs and even specialised representations such as haptic visualisations. 
Depending on the interface flavour, the appraiser must consider the specific features of the user interface, such as menus, buttons, drag-and-drop functionality, visual programming, etc.

Much like in the case of the environment, ergonomics and usability are important elements in terms of the suitability of the interface. 
For instance, a drag-and-drop command may produce an ergonomic visualisation interface --- it is easy to control, and quick to operate --- yet the number of available options may hinder the user's experience. Likewise, certain interactions in an immersive environment may be difficult to manage over time and the user get tired, or motion sickness affects them. 
Visual displays are often made from many different facets and coordinate views: are the sizes of these views right for the task? E.g., in a comparison task comparing data $A$ to data $B$, we may assume that the size given over to display both datasets is about the same (so not to bias one to the other). Or, in a webviewer, we may assume a central view will sit alongside adverts (are the size of the adverts suitable; perhaps too big or small in comparison to the visual display). Is spacing utilised efficiently, as it can enhance clarity and emphasise certain elements yet it can be detrimental as it can lead to confusion by disrupting desired relationships between elements.

\myPerspectiveIII{pictures/pic4-}
\noindent\textbf{Components:} Components are specific visual elements or depictions that can be identified and isolated for individual consideration. Identifying and understanding these components is essential for analysing the structure and effectiveness of the design/visualisation as a whole.
Each component serves a distinct purpose within the design, artefact or visualisation, contributing to the overall presentation and conveying specific information to the viewer.  Components can include various graphical elements such as charts, graphs, tables, icons, or other visual representations of data; an identified depiction~\cite{ChenETAL_2021}, along with menus, labels, help-information and so forth. They can be displayed in different ways, e.g., grid or in a tabbed window~\cite{al2019towards}.
The appraiser needs to consider if important components are missing, such as interactive elements, interaction handlers and buttons, or even appropriate data types.
Likewise, the type of output should facilitate interpretation and task execution, and align with the purpose of the presentation and the required task 
e.g., representing continuous data with a line graph and categories with a bar chart.
Considerations should also include whether the relationships between different parts of the display are evident, such as in the case of multiple view visualisation.

\myPerspectiveIII{pictures/pic5-}
\noindent\textbf{Design:} Design encompasses organising any part of the system, which involves considerations like colour balance, item alignment, and styling. This category focuses on more visually aesthetic aspects of the visualization or the supporting interface.
While attractiveness is subjective, humans tend to prefer balanced designs~\cite{Ware12}, those that are well-proportioned. Colour choices fitting with the task at hand are important~\cite{Harrower2013colorbrewer, BorlandTaylor2007}, whereas UIs following good design principles~\cite{DeiterRams, ShneidermanETALBookDesign} are always preferred. In terms of how the design facilitates sensemaking, there are questions regarding the representation, aggregation and appropriate (statistically) depiction of the data.
Finally, basic yet important elements such as legends, labels, titles, etc.~\cite{DykesETAL2010} can be of paramount significance, and only ommited if that makes sense. 

\myPerspectiveIII{pictures/pic6-}
\noindent\textbf{Visual marks:} The last step requires the appraiser to consider the graphical marks utilised in the visualisation and their arrangement, ensuring the appropriate marks are used in the correct locations with the correct attributes. Evaluating for redundant ink or chartjunk~\cite{tufte1983visual}, aspects of design~\cite{Ware12}, and how it remains memorable~\cite{borkin2013makes} are important considerations. Issues relating to scale, mark legibility, mapping validity and clarity for the display medium available are also important.
For multisensory visualisation systems, considerations beyond visual depictions, such tangible, vibrotactile and auditory representations require special considerations, including in terms of mapping efficiency, data transformation as well as accessibility~\cite{kim2021accessible}. Similarly, for immersive visualisations, issues such as occlusion should be also considered. 

\subsection{Stage~\texorpdfstring{\phaseTHREEitem}{3} -- Review}
The goal of this final stage (\cref{Fig:Stage3}) is to distil the key findings and observations, and translate these insights into actionable steps that contribute to an improved design/artefact. 
Initially a score is calculated by summing the Likert scale values 
and reflect on each part (the name, essence, six perspectives and so forth). Although the average score derived from the Likert scale calculation provides utility, it may be misinterpreted, as it conceals numerous facets of the actual critique. Therefore, it should be interpreted in conjunction with other insights. Review the six perspectives and the 30 questions. Identify any standout perspectives among the six. Assess the strengths and weaknesses, highlighting areas for improvement and identifying aspects to be strengthened.

After assessing the critique, it is important to consider the next steps. Perhaps a redesign to address identified issues and improve the overall design, which may involve brainstorming potential improvements, such as refining layout, focusing the adjusting visual elements, enhancing usability, or incorporating user feedback. The next steps would then include outlining a plan for redesign, which may include tasks such as conducting further research, gathering additional user input, creating prototypes, and implementing changes iteratively. It is essential to establish clear goals and objectives for the redesign and to regularly evaluate progress to ensure that the new design effectively addresses identified needs and enhances overall usability and user experience.

\begin{figure}[t]            
  \begin{tcolorbox}[title,width=\linewidth,
    left=0mm,
    middle=0mm,right=4mm,
    lower separated=true,
    lefthand ratio=.4,
    lefthand width=.2\linewidth,
     colback=\vignetteColour,colbacktitle=yellow!10!black,
     title={Stage~\numcircledtikz{3} -- Review},
     center title,
     before upper*=\fontsize{8}{9}\selectfont,
     before lower*=\fontsize{8}{9}\selectfont\begin{tabular}{@{}p{.5\linewidth}p{.5\linewidth}@{}},
     after lower*=\end{tabular},
   ]
      \vspace{-3mm}\textbf{Create score.}\hspace{1cm}
      \textbf{Reflect on parts:}\hfill
      \tcblower
       \textbf{Improvements and next steps:}&\\
     \end{tcolorbox}
\centering
        \vspace{-3mm}\caption{The final stage involves synthesising the various perspectives and insights, gathered throughout the critique.}\vspace{-3mm}
        \label{Fig:Stage3}
\end{figure}

\section{Design and Evolution of the CDS}
\label{SEC:DesignOfCDS}
\begin{figure*}[t]
	\centering
	\includegraphics[width=\textwidth]{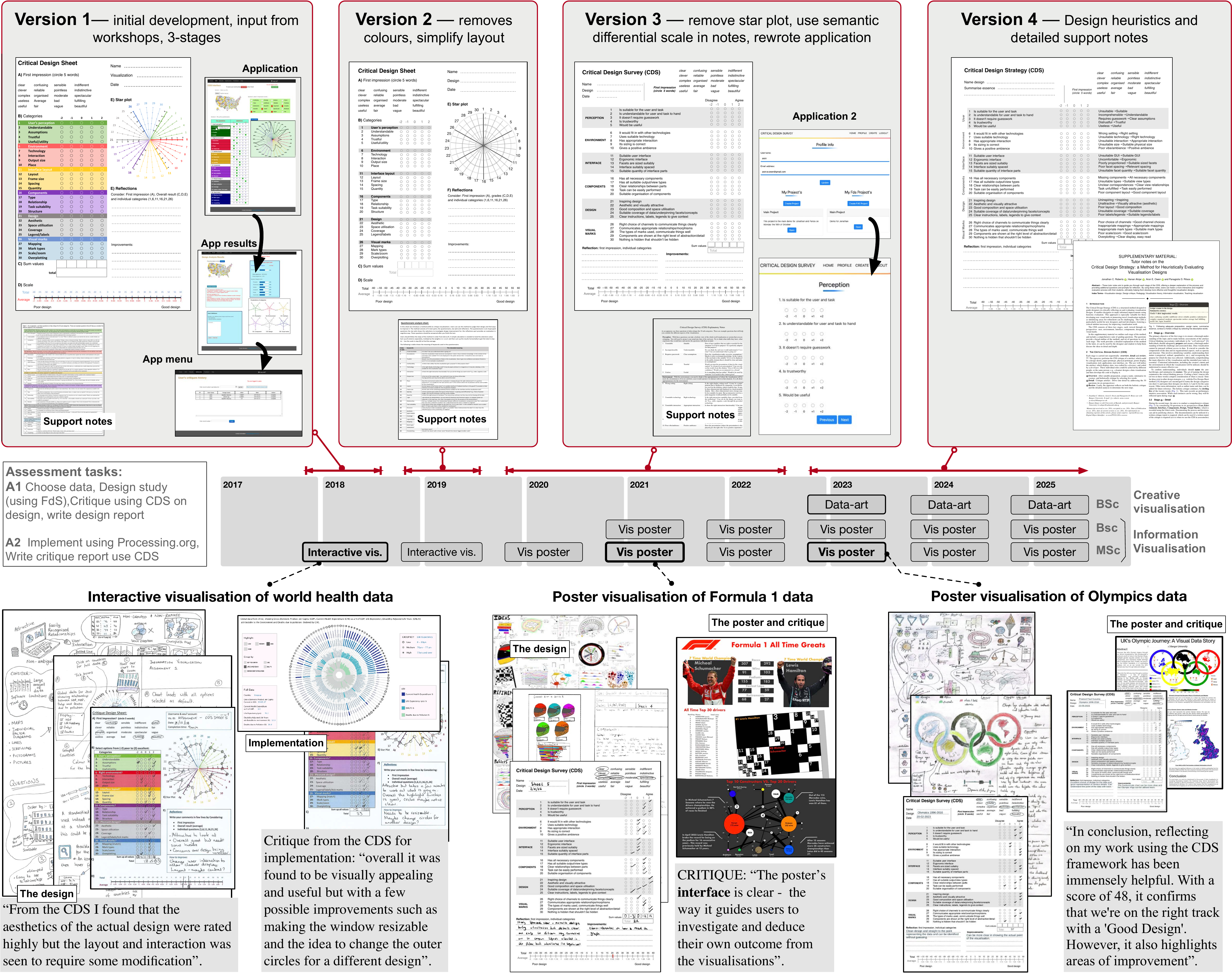}
	\caption{Timeline of CDS Versions, aligned with the courses where they were used (MSc or BSc in information Visualisation and Creative Visualisation) and the tasks (either build an interactive visualisation, or a static visualisation poster, or data-art piece). Three students' work are included, where they perform a design study then the CDS; then implement their work and perform another critique with the CDS. Quotes of their use of the CDS are included. Version 1 was developed from workshops, introducing the initial sheet, support notes, and web application while establishing the three primary stages, and six design perspectives. Version 2 refined the layout by removing colours and simplifying the sheet. Version 3 brought substantial changes, including the removal of the star plot and a complete rewrite of the application. Version 4 introduced semantic differential words (pairs of opposite adjectives to clarify design extremes) along with expanded and revised support notes. }
	\label{FIG:evolution}\vspace{-2mm}
\end{figure*}

In this section, we explore the design and evolution of the CDS over the past eight years (illustrated in \cref{FIG:evolution}), providing a critique of each version and explaining the rationale behind the development of its various forms. Throughout the design process, several methods were used to evaluate it, with the most significant being the application of the CDS in our teaching courses (see \cref{SEC:Background}), where it was tested in real-world conditions.

\subsection{CDS V.1 and V.2 2017-2019}
\label{SEC:Evaluation1}
\begin{figure}[t]
	\centering
	\includegraphics[width=.68\columnwidth]{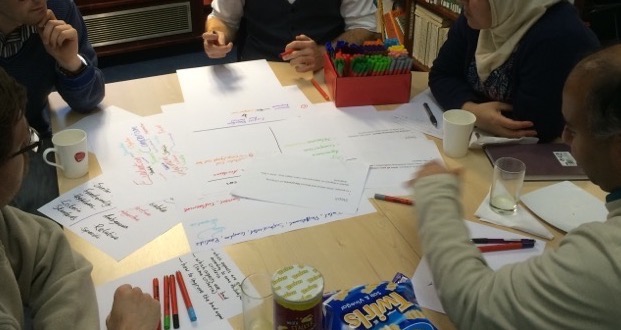}
    \includegraphics[width=.29\columnwidth]{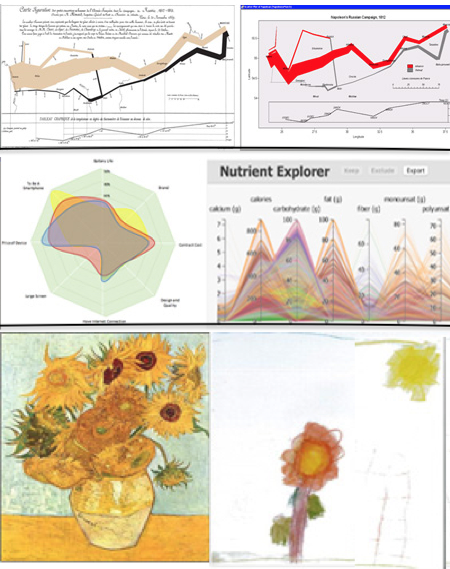}
	\caption{Photograph of participants at the critical-analysis workshop (left).
    Thumbnails of critiqued images (right). Minard's map of Napoleon's Russian campaign {\small(\copyright ~public domain)}; modern version of Minard's map {\small(DkEgy CC BY-SA 4.0}), radar plot, D3.js parallel coordinate plot. Reprint of Vincent Van Gogh's sunflowers, child's painting of same.}
	\label{FIG:WorkshopPhoto}
    \vspace{-5mm}
\end{figure}

The first version of the CDS was designed as a structured questionnaire.
We started with a 2-day workshop investigating critiquing vocabulary (see~\cref{FIG:WorkshopPhoto}), then designed the preliminary CDS structure and performed a talk-aloud evaluation~\cite{hanan2017}. We recruited 10 participants (six identified as male and four as female) with age between 25-40. Experiences ranged from ICT consultant, marketing expert, three doctoral students (visualisation, NLP and engineering), a postdoc in English, a mature student, with the remaining people were computing undergraduates. We formed two teams of five, and provided light lunch.
We asked participants to consider eight tasks: (\TN{1}) individually write a definition of critical analysis and underline key words in their definition; (\TN{2}) In groups, brainstorm over 15 words associated with critical analysis and produce a word-cloud. 
(\TN{3}) Critique the given six images (\cref{FIG:WorkshopPhoto}); (\TN{4}) Reflect on the critiques and consider if all critiques were the same. (\TN{5}) Write notes explaining similarities and/or differences. (\TN{6}) Think about and discuss the process of critiquing. (\TN{7}) Sketch a diagram to represent the main stages of the critiquing process (e.g., draw a flow diagram).
(\TN{8}) Using the same process critique in turn the parallel coordinate plot and radar plot visualisation. Two observers took notes, and we recorded conversations, which we later transcribed.

Analysis of the transcript and notes were informative, confirming ideas of critical thinking from Facione~\cite{Facione1990critical}. For instance, participants focused on the idea of serious examination and judgement; one wrote ``\textit{critical analysis is to perform in-depth thinking, for making fair and balanced judgements}''. Tasks T2,3,4 required teamwork, and to create a word cloud. 
When comparing Van Gogh's sunflowers to the kid's artwork (\TN{3}), \textit{``I guess this is basic and this advanced''} (pointing in turn to the pictures) and another said \textit{``one influenced the other''}. At that stage a participant added the word \textit{``influence''} to the word cloud, and promptly said: \textit{``we had some bias to some objects. We got influenced by our prior knowledge. We got influenced by thinking about the objects when we considered them''}; demonstrating awareness of bias~\cite{Facione1990critical}.  To analyse the vocabulary, we looked at the words that the candidates underlined (\TN{1}), word clouds (\TN{2}) and their discussions (\TN{3} and \TN{4}). Underlined words were predominantly verbs, e.g., \textit{think}, \textit{formulate}, \textit{imagine}, \textit{improve}, and \textit{analyse}. Participants used adjectives in written definitions, particularly with \TN{3}, and the word clouds. They include \textit{good}, \textit{bad}, \textit{useful}, \textit{clear}, and \textit{fair}. We summarise their words:
appealing,
	basic, 
	busy, 
	classic, 
	colourful, 
	confusing, 
	dull,
	fit for purpose, 
	fun,
	functional,
	influential,
	lack of context,
	minimalist,
	modern,
	old fashioned, 
	plain, 
	pretty, 
	realistic perspective,
	reusable,
	simple, 
	sophisticated, realistic scale,
	surreal,
	too detailed, 
	unreadable.

When discussing the similarities and differences (\TN{5}) team~1 distilled their ideas into four categories, expressing that these cross-cut all the object and image types: (1) the \textit{style} (and layout) of an image, (2) its \textit{appearance} and what it looks like, (3) \textit{comparison} with other ideas and to an idea of what it could look like, and (4) whether it was suitable for the \textit{context} (or environment) where it would be used. Participants acknowledged that they needed to put their own emotions aside,  become more `academic' in their approach, and that some objects/pictures engendered more emotional responses than others. If they understood the image and its purpose, or if it was something they had seen, liked or used before, then their emotional response was stronger. They also emphasised that to critique the objects each participant needed to understand the \textit{context} where it would be used. These responses match similar judgements by Ennis~\cite{ennis2011critical}, where users need to employ \textit{logic} to understand the judgement \textit{criteria} and be \textit{pragmatic} in their approach.  For \TN{5}, team~2 likewise included \textit{appearance} and \textit{comparison}, but also emphasised the importance of the \textit{initial thoughts} to the design, and in comparison to deep thought. This result is significant, and implement it by the action to `circle 5 words' of the list of 20. 

Finally, addressing the last three tasks (\TN{6},\TN{7} and \TN{8}) we focused on team~2, as they completed the tasks in more detail and their answers are a superset of what group~1 produced. They produced a useful four-stage process,
where the observer (1) thinks about the problem and defines the terms of critical analysis, (2) brainstorms ideas and organise their thoughts, (3) produces a balanced review by putting emotions aside, and finally (4) reflects on the analysis to reject or accept the critique. This is another important result because it confirms the process described by Polya~\cite{Polya2014solve} on problem-solving, where users should understand the problem, devise a plan, carry it out, and confirm the answer.

\subsection{Developing the CDS strategy (V1)}
\label{SEC:Evaluation2}
To develop a list of suitable `first-impression' words, we started with the list of adjectives. We used card-sorting, to decide on a list of 20 words. We made a compromise between having too many words and confusing the user, to not having enough words that could inhibit expression. We wanted to balance positive, neutral and negative sentiment words, and for that we used scores from the \textsc{Sentiwordnet}~3.0 lexicon~\cite{Baccianella2010}. Sentiment analysis depends on the position of the word and how it is used, and each of our words have multiple sentiment scores in the lexicon~\cite{Taboada2011}. We placed all words in a table and recorded an average of all scores in the lexicon for each word. After discussion we simplified the list to seven 
positive (\textbf{average, beautiful, clear, clever, reliable, sensible, spectacular}), 
seven negative (\textbf{bad, complex, indistinctive, pointless, confusing, useless, vague}) 
and six neutral (\textbf{fair, fulfilling, indifferent, moderate, organised, useful}) words.

For our first design, we tried to follow the System Usability Scale (SUS)~\cite{bangor2008}, and alternated between positive and negative wording of questions. The advantage of this strategy is that it combats response acquiescence, minimising the results from participants who (say) make quick, injudicious, selections on one side of the Likert scale. We developed a list of questions, including ``\textit{I found the colour map suitable''.}
\textit{``I would imagine that most non-expert users would not easily understand data that is presented by this visualisation without any prior knowledge of the visualisation domain''.} \textit{``My first impression is that I have a high confidence in this visualisation''.}
\textit{``I felt the visualisation has not achieved the design goals.''}
However, we faced several challenges. First, it was difficult to distil the subject to ten questions, as we did not believe these were enough for us to perform an in-depth critique of a visualisation design. Second, the questions seemed too convergent; we wanted the user to really think deeply about their work, not merely judge it whether it was right or wrong. Third, when we tested these questions, we found it difficult to complete the responses of alternating positive to negative question wording. We discussed this problem with colleagues, who had not been involved in our design or workshop, who likewise found it difficult to use. 

Reflecting on this issue we realised that the goal of the SUS is different, as it is designed for summative assessment from end-users, with the results analysed once a prototype has been finalized. However, we want the designer/developer to use our method for their own work in a formative manner and make decisions on their own creations. For our situation, there is no need to alternate positive and negative to combat response acquiescence -- why would a user not be honest with themselves -- and the questions were too convergent to achieve our goals of prompting the user to think critically about the problem.  
We therefore looked to other questionnaires.

The UEQ~\cite{Rogers2011} measures user-experience through a set of six categories: \textit{Attractiveness}, \textit{Efficiency}, \textit{Perspicuity}, \textit{Dependability}, \textit{Stimulation} and \textit{Novelty}, gauged using a 5-part Likert scale of (alternating positive and negative) 26 questions. The UEQ takes longer to complete than the SUS, and the creators used statistical analysis to develop a set of weights, to balance the results to an `average user'~\cite{Laugwitz2008}.  However this further complicates how the results are calculated. We did not want users to have to input their values into a spreadsheet, so decided to implement a quick way to calculate the scores. Weighing up the different strategies, we decided to follow the UEQ design to develop a list of 30 questions. 

To develop the six perspectives of the CDS, we were heavily influenced by the words of one workshop participant (drawing on Shneiderman's mantra~\cite{shneiderman1996eyes}) who said  \textit{``when we are critiquing we need to be clear what we are looking at, we first need to understand what it is, then put aside bias and emotion. There are different levels of detail. We should look to the big picture before looking at details''}. We can consider design critique in a similar way. Start by thinking holistically and drill down into the finer detail. People can critique a design by thinking about the \textit{user}, then the usage \textit{environment}, the \textit{interface} that would be used in said environment, and so on, until we get to individual graphical marks that are used to create the visualisation.

The next task was to develop the individual questions of the framework. Starting with key terms from the workshop we used a card-sorting process to categorise them into the six perspectives. We show these terms in \cref{tab:30words-version-1}. Placing the cards on a table, as a team, we moved cards with similar meaning words on top of each other, with the goal to reduce the overall quantity of indicative words to five per category. (Five cards, in six perspectives, gives 30 questions, which is close to 26 topics in the UEQ). 
We deliberated over individual parts, for instance whether we should include \textit{`design'} or more specific terms, such as \textit{`colour'}. 
From these prompt words we added words with the opposite meaning, and placed them on a Likert scale; 0 for a poor value (left), and 5 a good value (on the right). This meant that we could add up the values and calculate an overall score, without requiring to compensate for alternating scales (as in the SUS).

We deliberated much about creating an overall score of the user's critique. On one hand, having a single value that can be awarded to a visualisation is useful. For example, if a user uses the CDS for a second time, the overall score difference would allow them to quickly decide whether the visualisation has improved. On the other hand, a single average score, could potentially mask bad perspectives.  
This is similar to Anscombe's quartet~\cite{anscombe1973graphs} of statistical graphs, that look completely different but have similar statistical properties. We deliberated over using six scores (one for each category), and even considered not including a score at all. We decided to keep the score as we believe the benefits outweigh the negatives.

We evaluated this first version with three talk-aloud evaluation sessions (S1, S2, S3). In S1 we focused on comprehension of the initial thirty `prompt' words (\cref{tab:30words-version-1}); S2 and S3 focusing on details of the questions and on the Likert scale and overall score. For each talk-aloud 90-minute session we used two researchers (one with expertise in information visualisation and the other in virtual reality). We explained the heuristics, they read the explanatory notes, and we video recorded the sessions and transcribed their comments. We used six different visualisations; two of each: line-graph, bar chart, parallel coordinate plot. Our goals were to ascertain if the strategy was easy, quick to complete and comprehensive. 
We made several changes from these talk-aloud evaluations. First, we adjusted some terms. E.g., (\#10) \textit{place} and \textit{`ambience'}, and (\#11) \textit{interface} to \textit{GUI}; \Cref{tab:30words-version-1} shows the \st{old} and new words. Second, participants said the 5-part Likert was useful, but took too long to calculate the result. They suggested to swap it from $[1 ...5]$ to $ -2$ to $+2$, with $0$ being the mid-point. It is now quicker to sum the values, and the calculation gives an idea that the design was poor or good. Third, participants recommended adding a star plot to summarise the 30 scores. We implemented this in Version 1 and used it for the first two years, but removed it from Version 2 onwards due to the time it took to complete manually. 

\setlength{\belowcaptionskip}{0pt}

\begin{table}[t]
	\def\arraystretch{.98}
    \centering
        \caption{The terms used in our card-sort exercise, for the 30 questions. Old words (\st{striked}), and new words after the talk-aloud exercise.}\vspace{-2mm}
    \label{tab:30words-version-1}
    \fontsize{7.2}{8}\selectfont
    \begin{tabular}{@{}l@{\hspace{1mm}}l@{\hspace{2mm}}l@{\hspace{1mm}}l@{\hspace{2mm}}l@{\hspace{1mm}}l}
\#1&\st{Perception}Suitable&\#11&\st{Interface}GUI&\#21&\st{Design}Inspiring\\
\#2&Understandable&\#12&\st{Layout}Ergonomic&\#22&Aesthetic\\
\#3&Assumptions&\#13&Frame size&\#23&\st{Space utilisation}Composition\\
\#4&Trustful&\#14&Spacing&\#24&Coverage\\
\#5&Useful/utility&\#15&Quantity&\#25&Legend/labels\\
\#6&\st{Environment}Setting&\#16&Components&\#26&\st{Visual marks}Channels\\
\#7&Technology&\#17&Type&\#27&Mapping\\
\#8&Interaction&\#18&Relationship&\#28&Mark types\\
\#9&Output size&\#19&Task suitability&\#29&Scale/zoom\\
\#10&\st{Place}Ambience&\#20&\st{Structure}Layout&\#30&Overplotting\\
\vspace{-7mm}
    \end{tabular}

\end{table}

\subsection{Analysis of Version 2, and developing V3 (2019-2023)} 
At this time the cohort size doubled, and the information visualisation module was incorporated into a wider range of programmes (data science and a general computing programme), attracting students with less advanced programming skills. As a result, we transitioned from having students develop an interactive visualisation of selected data (2018-2019) to designing a new visualisation for a poster display. This shift removed interaction from the coding process and placed a greater emphasis on design and storytelling (see \cref{FIG:evolution}). We continued to guide students in focusing their critical thinking on the three stages (\texorpdfstring{\phaseONEitem}{1} overview, \texorpdfstring{\phaseTWOitem}{1} detail, and \texorpdfstring{\phaseTHREEitem}{1} review). Our lecture approach remained similar, covering topics such as visualisation history, design, and perception. In addition to a design lecture on the FDS method, we delivered an hour-long lecture on general critical thinking, followed by an explanation of the CDS structure. We maintained the two-assessment tasks: a design report and an implementation using Processing. Students were asked to complete a CDS sheet at least twice (once for each assessment), either by hand or digitally, and include a copy in their report. Most chose to complete it by hand and submitted a scan with their report.

By this point, we had been using the CDS in our teaching for two years, but we aimed to evaluate Version 2 more formally. To do so, we conducted two separate usability evaluation sessions with different student groups Group~1 (\GN{1}) had 10 PhD students from the Computer Science department; half had experience in creating visualisations with JavaScript and all had created charts in Excel. Group~2 (\GN{2}) were 20 BSc computer science major students, who were all taking the visualisation module and using \href{http://processing.org}{processing.org} to create visualisations. All participants were volunteers and none received compensation for participating. 
We gave a 15 minute introduction, consent form, printed CDS, explanation text of 30 categories, and six visualisation scenarios. They had an hour to reflect on each scenario and judge the visualisations. We chose the visualisations to include familiar and (potentially) unfamiliar forms. These were (i) a GapMinder bubble chart, (ii) an Excel pie chart, (iii) a treemap, (iv) a transport network map, (v) a time series graph, and (vi) an Excel bar chart.  The written scenarios placed the visualisations in a particular context. E.g., the pie chart visualisation showed University annual research-grant income, and was to be included in the University's research report.  
Each participant was observed, and notes and timings recorded. Participants were asked to critique the visualisations in order. 
In \GN{1} eight participants completed six, wheras two completing five scenarios.  In \GN{2}, only one participant did not finish within the hour. We note, that students in \GN{2} used the CDS three additional times, as they applied it in their teaching while submitting assessments 1 and 2.

\setlength{\arrayrulewidth}{.1em}
\begin{table}
	\small\sffamily
	\renewcommand{\arraystretch}{1.1}
 \fontsize{7.}{8}\selectfont
	\caption{The internal-correlation of 30 users (N=30) users completing the categories on six different visualisations. Time taken for \GN{2} (N=20).\vspace{-2mm}}
	\label{alpha-table}
	\begin{tabular} {llll}
		\hline
		\textbf{Visualisations}  & \textbf{Cronbach's}  \scalebox{1.}{$\alpha$} & \textbf{Variance}&\textbf{Group2 time}, (min)\\  
		Bubble chart & .89 & 0.080 &15.6 \\ 
		Pie chart  & .91 & 0.157& ~~9.8\\ 
		Treemap chart & .92 & 0.130& ~~8.4 \\ 
		Network graph & .95 & 0.041 &~~8.0\\ 
		Time series graph & .92 & 0.068 &~~7.0 \\ 
		Bar chart & .95  & 0.057 &~~6.3\\ 
		\hline
	\end{tabular}
\end{table}

\begin{table}[t!]
	\small\sffamily
	\renewcommand{\arraystretch}{1.1} 
	 \fontsize{7.}{8}\selectfont
	\caption{Time taken for each visualisation. Independent samples t-test results show $\rho$-value at significance level a=0.05, associative mean and SD for each visualisation. \vspace{-2mm}
	}
	\label{sigp-table}
	\begin{tabular} {lcrl m{2.5cm}}
		\hline\\[-2.5mm]
		\textbf{Visualisations}  & \textbf{Group} &\textbf{Mean} & \textbf{SD} &\textbf{ Sig. $\rho>$0.05 } \\ [1mm]
		Bubble chart & 1 & 102.6 & 18.4 &  .564 \\
		& 2 &108.0 & 13.2 & \\ 
		Pie chart  & 1 & 113.9& 20.4& .623 \\
		& 2& 99.2 & 16.13   &\\
		Treemap chart &1 & 78.6 & 21.5& .728\\
		& 2 &67.9 &17.4 &\\ 
		Network graph & 1 & 111.9 & 11.4& .007$^*$\\
		& 2& 90.6 &25.17 &\\ 
		Time series graph & 1 & 105.2 & 11.2& .051  \\
		& 2& 112.8&19.0 &\\ 
		Bar chart & 1  & 102.8 &19.2& .53\\
		& 2 & 122.0 & 17.4 & \\
		\hline
	\end{tabular}
\vspace{-4mm}
\end{table}

To analyse the inter-item consistency (across PhD and undergraduate participants) we first considered all participants together (N=30). 
Cronbach's alpha values (\cref{alpha-table}) indicates that users did broadly answer in a similar, consistent way. Putting aside potential issues with Cronbach's $\alpha $~\cite{Tavakol2011} where the calculation can be sensitive to the number of scales, we believe this is a positive result.
To investigate the difference between \GN{1} (PhD) and \GN{2} (BSc) we used a t-test and made the dependent variable the total score of each visualisation. \cref{sigp-table} shows the calculated significance $\rho$ values for each visualisation. The $\rho$-value for the bubble, pie chart, treemap and bar chart visualisation are greater than the chosen significance level 
${a} =0.05$, so we don't reject ${H_0}$ the null hypothesis that there is no statistically significant difference (variance) of the means between users groups. However, the null hypothesis for equal variance can be rejected for the network map because  $\rho$-value is less than $0.05$. That means there is statistically significant difference for the mean values between groups. For timeline chart, $\rho$-value is equal to $0.05$, yet the null hypothesis is rejected based on the variance between groups. In addition the associative SD for visualisations that has $\rho$-value $< 0.05$ have big mean difference between values (more than 4) from the average values to other visualisations. 

\definecolor{RYB1}{RGB}{252,96,96}
\definecolor{RYB2}{RGB}{255,149,53}
\definecolor{RYB3}{RGB}{255,186,80}
\definecolor{RYB4}{RGB}{189,229,146}
\definecolor{RYB5}{RGB}{127,229,185}
\definecolor{RYB6}{RGB}{69,117,180}

\pgfplotsset{
	tick label style = {font=\sffamily},
	every axis label = {font=\sffamily},
	legend style = {font=\sffamily},
	label style = {font=\sffamily}
}

\Cref{fig:imp1} shows first-impression word use, across both groups. While there are some differences between groups, there is overall consensus. Words like \textit{confusing}, \textit{complex}, and \textit{bad} were more frequently chosen, while terms like \textit{spectacular} and \textit{fulfilling} were less common. Since participants in \GN{1} engaged in more discussion, we focus on completion times for \GN{2}, shown in \cref{alpha-table}. Participants became faster with each use, reaching completion times of around 5 to 8 minutes.
Finally, participants commented on positive and negative aspects of the CDS, and how the CDS helped them to critique their own visualisations. Twelve participants replied, saying \textit{``it helps you see on paper what's good and what's bad''}. \textit{``The CDS asks you to be honest and rate different properties of the program, which helps spot things that could be improved that would have been overlooked without CDS''}. Another wrote \textit{``visually very easy to see how the design scored, and gives developers clear areas in which they can improve''}.

\setlength{\belowcaptionskip}{-5pt}
\begin{figure}[t]
	\centering 
	\tiny
\resizebox{.7\columnwidth}{!}{
\begin{tikzpicture}
\begin{axis}[
grid=major,
grid style={line width=.1pt, draw=gray!10},
point meta=explicit,
xmin=-1,
xmax=12,
ymin=Bad,
ymax=Spectacular,
enlarge y limits={0.04},
symbolic y coords={Bad,Complex,Indistinctive,Pointless,Confusing,Useless,Vague,, Fair, Fulfilling, Indifferent, Moderate, Organised, Useful,, Average, Beautiful, Clear, Clever, Reliable, Sensible, Spectacular},
ytick = {Bad,Complex,Indistinctive,Pointless,Confusing,Useless,Vague,, Fair, Fulfilling, Indifferent, Moderate, Organised, Useful,, Average, Beautiful, Clear, Clever, Reliable, Sensible, Spectacular},
y dir=reverse,
xtick = {0,...,11},
xticklabels={G1,G2 ,G1,G2 ,G1,G2 ,G1,G2 ,G1,G2 ,G1,G2 ,G1,G2 ,G1,G2},
x=0.4cm,
y=0.35cm,
xtick style={draw=none},
ytick style={draw=none},
extra x ticks={.5, 2.5, 4.5, 6.5, 8.5, 10.5},
extra x tick labels={Bubble, Pie, Treemap, Network, Time, Barchart},
extra x tick style={
	ticklabel style={yshift=-7pt, font=\bfseries\sffamily\tiny},
	tickwidth=0
},
extra y ticks={Indistinctive, Indifferent, Clever},
extra y tick labels={Negative, Neutral, Positive},
extra y tick style={
	ticklabel style={xshift=-30pt, font=\bfseries\sffamily\small, rotate=-270},
	tickwidth=2
}
]
\addplot[only marks, scatter, scatter src=explicit, mark=*, 
scatter/@pre marker code/.code={%
	\pgfmathparse{\pgfplotspointmetatransformed/1200*100+50}%
	\let\myopacity=\pgfmathresult
	\pgfmathparse{\pgfplotspointmetatransformed/1200*7.5}%
	\def\markopts{mark=*, color=RYB1, fill=RYB1!\myopacity, mark size=\pgfmathresult}%
	\expandafter\scope\expandafter[\markopts]
},
scatter/@post marker code/.code={\endscope},
] table[x index=0, y index=2, meta index=3] {data/bubble.dat};
\addplot[only marks, scatter, scatter src=explicit, mark=*, 
scatter/@pre marker code/.code={%
	\pgfmathparse{\pgfplotspointmetatransformed/1200*100+50}%
	\let\myopacity=\pgfmathresult
	\pgfmathparse{\pgfplotspointmetatransformed/1200*7.5}%
	\def\markopts{mark=*, color=RYB2, fill=RYB2!\myopacity, mark size=\pgfmathresult}%
	\expandafter\scope\expandafter[\markopts]
},
scatter/@post marker code/.code={\endscope},
] table[x index=0, y index=2, meta index=3,] {data/pie.dat};
\addplot[only marks, scatter, scatter src=explicit, mark=*,
scatter/@pre marker code/.code={%
	\pgfmathparse{\pgfplotspointmetatransformed/1200*100+50}%
	\let\myopacity=\pgfmathresult
	\pgfmathparse{\pgfplotspointmetatransformed/1200*7.5}%
	\def\markopts{mark=*, color=RYB3, fill=RYB3!\myopacity,  mark size=\pgfmathresult}%
	\expandafter\scope\expandafter[\markopts]
},
scatter/@post marker code/.code={\endscope},
] table[x index=0, y index=2, meta index=3] {data/treemap.dat};
\addplot[only marks, scatter, scatter src=explicit, mark=*, 
scatter/@pre marker code/.code={%
	\pgfmathparse{\pgfplotspointmetatransformed/1200*100+50}%
	\let\myopacity=\pgfmathresult
	\pgfmathparse{\pgfplotspointmetatransformed/1200*7.5}%
	\def\markopts{mark=*, color=RYB4, fill=RYB4!\myopacity, mark size=\pgfmathresult}%
	\expandafter\scope\expandafter[\markopts]
},
scatter/@post marker code/.code={\endscope},
] table[x index=0, y index=2, meta index=3,] {data/network.dat};
\addplot[only marks, scatter, scatter src=explicit, mark=*,
scatter/@pre marker code/.code={%
	\pgfmathparse{\pgfplotspointmetatransformed/1200*100+50}%
	\let\myopacity=\pgfmathresult
	\pgfmathparse{\pgfplotspointmetatransformed/1200*7.5}%
	\def\markopts{mark=*, color=RYB5, fill=RYB5!\myopacity,  mark size=\pgfmathresult}%
	\expandafter\scope\expandafter[\markopts]
},
scatter/@post marker code/.code={\endscope},
] table[x index=0, y index=2, meta index=3] {data/time.dat};
\addplot[only marks, scatter, scatter src=explicit, mark=*, 
scatter/@pre marker code/.code={%
	\pgfmathparse{\pgfplotspointmetatransformed/1200*100+50}%
	\let\myopacity=\pgfmathresult
	\pgfmathparse{\pgfplotspointmetatransformed/1200*7.5}%
	\def\markopts{mark=*, color=RYB6, fill=RYB6!\myopacity, mark size=\pgfmathresult}%
	\expandafter\scope\expandafter[\markopts]
},
scatter/@post marker code/.code={\endscope},
] table[x index=0, y index=2, meta index=3] {data/barchart.dat};
\end{axis}
\end{tikzpicture}
}
	\caption{Analysis of the 20 first-impression words. The plot demonstrates a broad use of most words, apart from spectacular, fulfilling and beautiful.} 
	\label{fig:imp1}
	\vspace{-2mm}
\end{figure}

After analysis, we made improvements to the CDS. We removed the star plot, as it was time-consuming and not effectively completed. Participants (across \GN{1} and \GN{2}) requested clarification of some terms, possibly due to students not fully reading the explanatory descriptions. These changes are outlined in \cref{tab:30words-version-1}. Based on feedback and the words in the supplementary material, we introduced the semantic differential scale, inspired by the UEQ-style questionnaire. While we continued to emphasise the heuristic questions in the taught classes and notes, the semantic differential words helped quickly clarify the idea behind each of the 30 heuristics. Additionally, we revised the lecture material to carefully explain each heuristic and expanded the supplementary notes. This led to Version 3, in which we removed the star plot and incorporated the semantic differential scale in the notes. We also rewrote the application to align with this version, and to allow users to review both current and past CDS evaluations (see \cref{FIG:evolution}).

\subsection{CDS V.4 2023-present and its use} 
In our current practice, when integrating the CDS (as in ~\cref{SEC:TheCDS}) into assessments, we start with lectures on critical thinking and the CDS structure, and provide detailed notes (included in the supplementary material). Students then participate in a group activity~\cite{roberts2022reflections}, first collecting and discussing good and bad visualisations on a shared virtual whiteboard, then applying the CDS to two visualisations (one poor, one good). These exercises build critical thinking skills and with the CDS. Critique is inherently challenging and time-consuming to learn~\cite{Facione1990critical} (see \cref{alpha-table}), but the three-stage structure (overview, detail, review) and heuristic questions make it more manageable. With practice, efficiency improves.
Similar to other years, students design a solution, critique it using the CDS, and submit a report with a scanned CDS. For the second assessment, they develop their tool in Processing and apply the CDS again. Writing the report enhances focus and deepens analysis.

\section{Discussion and conclusions}
\label{SEC:DiscussionConclusions}
We asked the recent cohort (the BSc and MSc in information visualisation) about their experience, receiving 18 replies. The CDS framework significantly shaped critical thinking by providing a structured approach. One respondent said, ``\textit{It forced me to think about the nuances for my poster}''; another said ``\textit{The framework significantly enhanced my critical thinking by providing a structured approach to analyse my design}''. The 30 sub-questions were mostly helpful, with one person noting, ``\textit{It became useful to find any shortcomings}''. However, some found some heuristics less relevant, ``Some sub-questions felt less useful, especially not relevant to my design goals''. The first-impression words played a key role in shaping reflections. One participant said, ``\textit{It helped me streamline the presentation}'' and another felt, ``\textit{It inspired me to think from a fresh perspective}''. The ``User'' category was the most useful for many, with one saying, ``\textit{User environment, it made me think of the design from a different perspective}''. ``Components'' and ``Visual Marks'' were also valued for tracking progress and evaluating visual elements. As one person put it, ``\textit{Components helped me keep track of my progress}''. Some suggested changes, like simplifying questions. One respondent said, ``\textit{30 questions feels slightly too many'}'. Others suggested adding categories like ``\textit{Feedback \& Iteration}'' or ``\textit{Emotional Impact}'' to assess user sentiment. Most participants plan to use the framework in the future. One said, ``\textit{Yes, I will continue using the framework}'', while another noted, ``\textit{I would use it in my project report}''. However, one wrote ``\textit{I would not continue using the framework in future design critiques because it feels somewhat rigid'}'. We recognise that the process takes time to complete, and is detailed, but this is intentional, as it encourages a thorough and in-depth critique of the design.  Overall, the CDS framework was seen as valuable for critical thinking and design critique. As one person concluded, ``\textit{It was a helpful tool for reflection and improving my design process}''.

After eight years of using the CDS in various contexts, we have gained valuable insights and identified key lessons. \textbf{Vocabulary and confidence are essential.} Many individuals struggle to critique because they don't have the vocabulary. To address this, we provide (i) a lecture on critical thinking, and emphasise specific vocabulary, particularly phrasing from Facione~\cite{Facione1990critical}, (ii) class activities that provide a safe space to practice, and (iii) use repetition (requiring students to apply the CDS at least twice, once for each assessment). 
\textbf{Preparation is essential.} To effectively critique a visualisation, it is crucial to understand the data, purpose, and context. Without this foundation, evaluating its suitability becomes difficult. We support this by encouraging learners to consider the five W’s 
and to imagine the intended user. A clear articulation of the design scenario and requirements by the instructor helps set the context, and when possible, having a client present a real-world problem adds valuable depth. Additionally, the ``assign a name'' task (Stage~\texorpdfstring{\phaseONEitem}{1}) sharpens focus, by creating a concise title learners can better identify the core purpose, offering a clear reference point for discussion. 

\textbf{Emphasise the process, not the answer.} Some appraisers may merely want a score to grade the design, while others want to write a fast critique. We emphasise to our students that they must spend time on the exercise and think carefully. The score is only one aspect of the strategy, and that the strategy and \#30 questions are designed to prompt a deep consideration of the design. 
We emphasise that the CDS is a tool, to work alongside current design methods. It helps appraisers critique a variety of outputs. We do not want people to worry that they are not putting the right value on the form, rather, the CDS should be viewed as strategy to consider pros and cons with the visual design. 

\textbf{Write a report.} Needing to write a report helps students focus on the process rather than the result. From our evaluation, we know this is occurring: e.g., one student wrote ``\textit{The CDS helped me with the final tool, and where I was going wrong on Sheet5 of the FdS design}''. And another, when reflecting on a visualisation poster wrote ``\textit{the Design. Visual marks, Environment proved to be the lowest scoring sections, each with 6 points}'',  consequently they decided to change the colour palette, saying ``\textit{a more neutral palette may aid in rectifying this ... and more annotations}''. Another wrote,
``\textit{the CDS was useful to show myself how much of a difference there is between the goal I set out to achieve and what I actually managed to do.}'' and another ``\textit{ 
It provided a very good structured way for me to critique ... [and consider] 
points I would not have thought of.}'' 
\textbf{Adopt a broad-minded approach.} We encourage students to interpret the heuristics generally and apply them to their problem. E.g., question \#8 about the appropriateness for the interaction, could be interpreted if it has WIMP controls; but should be generally applied to consider the concept of `interaction' broadly. 

The CDS is a three-stage critical thinking framework (overview, detail, review) with 30 core questions designed to help individuals critique their designs and visualisation tools. Our strategy fosters critical thinking skills, which are essential for making informed decisions, by guiding appraisers in evaluating strengths, weaknesses, and areas for improvement. It provides a structured approach to deep thinking. To be most effective, the process should be given ample time, allowing for careful consideration of each aspect. We have used it in critical-thinking assessments, in both the design and implementation stages. Yet, 
it can be applied in different situations, and from several roles, including learners, researchers, and visualisation developers. We believe it is a valuable tool for tutors in teaching and assessing critical thinking. Ultimately, it empowers designers to make honest critiques and deeply reflect on their designs to create effective, engaging visualisation systems.

\clearpage
\bibliographystyle{abbrv-doi-hyperref}

\section*{Supplemental Materials}
\label{sec:supplemental_materials}
In the supplemental materials we include the full CDS explanatory notes, the final versions (with and without the differential scale) as resources for teaching.

\acknowledgments{
We acknowledge the help of the student learners across the years. Their substantial feedback (both positive and negative) has helped to improve and refine the CDS. We acknowledge the initial workshop participants, which ran as an evaluation as contribution for Alnjar's PhD\cite{hanan2017}; while the latter evaluations received ethics clearance CSEE-2021-CG-001 and CSE-2025-0678.}
\balance 
\bibliography{CDS_bibliographyNew}

\clearpage
\appendix

\section*{APPENDIX 1: SUPPLEMENTAL MATERIAL:
}
\textbf{Support notes on the 
Critical Design Strategy: a Method for Heuristically Evaluating Visualisation Designs 
}

\section*{Abstract}
These notes offer a detailed explanation of each stage of the Critical Design Strategy (CDS), including additional questions and prompts for reflection. They serve as a resource for tutors integrating the CDS into their teaching and for appraisers applying it in their evaluations. By providing commentary on each stage, these notes support both tutors and appraisers in fostering more effective and thoughtful visualisation designs.

\setcounter{section}{0}
\section{Introduction}

The Critical Design Strategy (CDS) is a structured framework that guides designers in critically evaluating and refining visualisation designs through heuristic assessment. It is especially valuable for those creating new visual tools or experimenting with innovative visualisation methods, as it fosters critical thinking about the design. 

The CDS consists of three key stages (overview, detail and review), with the detail section covering the critique through six perspectives: user, environment, interface, components, design, and visual marks. 

These notes are intended to support the appraiser by providing a detailed breakdown of each stage, along with additional questions to facilitate a thorough critique. We strongly emphasise the importance of deeply engaging with the process. It should go beyond a quick glance at the design, requiring a deliberate and thorough evaluation instead. Only by actively questioning and critically analysing the work can we fully understand its purpose, anticipate its use, and determine whether the design is truly effective and suitable.

\begin{figure}[h]
    \centering
\includegraphics[width=\columnwidth]{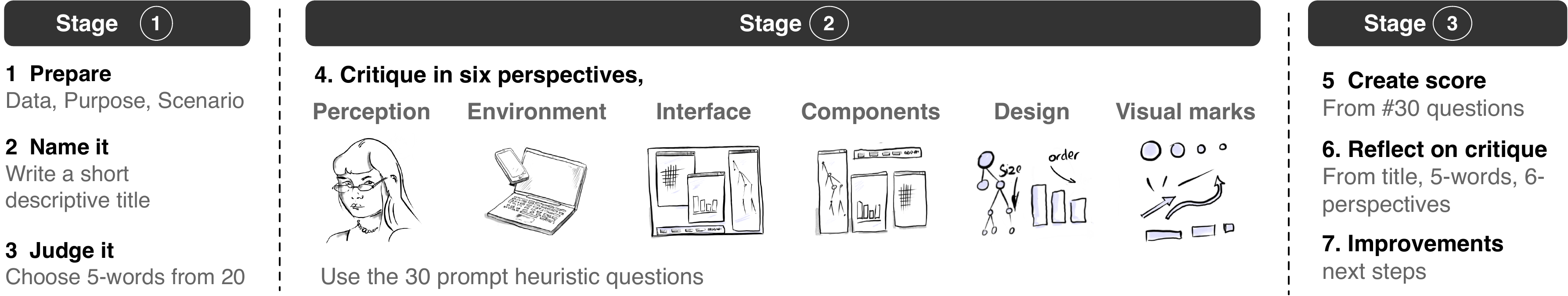} 
    \label{fig:enter-label}
    \caption{The Critical Design Strategy consists of three stages: \texorpdfstring{\phaseONEitem}{1} overview, \texorpdfstring{\phaseTWOitem}{1} detail, and \texorpdfstring{\phaseTHREEitem}{1} review. The assessor begins by considering the design holistically—naming and summarising it while selecting five keywords from a set of twenty. Next, they conduct an in-depth evaluation across six perspectives, using 30 heuristic questions or directly engaging with semantic differential word pairs (opposite adjectives). Finally, the assessor reflects on their critique, assigns an indicative score, and determines areas for improvement.}
    \label{fig:PartsOfThePaperFIG2}
\end{figure}
\renewcommand{\myLikert}{$\circ\circ\circ\circ\circ$}

\section{The CDS: Summary and Teaching Scenario}
\label{AppendixSEC:TheCDS}
Each stage is carried out sequentially: \textbf{overview}, \textbf{detail} and \textbf{review}, \cref{fig:PartsOfThePaperFIG2}. The \textit{appraiser} performs the CDS critique of a artefact, which could be a design sketch, paper prototype, physical prototype, poster display, visualisation, tool, application, user interface, etc. The \textit{user} will utilise the artefact, which displays \textit{data}, was crafted by a \textit{designer}, and coded by a \textit{developer}. These individual roles could be achieved by different people, or the same person; e.g., a learner designs a data visualisation and then develops the code to display it.

To achieve the CDS critique, follow the three stages in turn. To start the appraiser must understand the data and situation of where/how the artefact will be used. They need to prepared will be used, and understanding the data, scenario and how the visualisation will be used, then think what is important in the data (the design essence), and how to summarise the idea.
\vspace{1mm}
\begin{itemize}[nosep,leftmargin=\parindent, align=left, labelwidth=\parindent, labelsep=0pt]
    \item[\texorpdfstring{\phaseONEitem}{1}] {\textbf{Overview}}. After suitable preparation,
    assign a name, summarise its essence, and holistically critique by selecting five words.
    \item[\texorpdfstring{\phaseTWOitem}{2}] {\textbf{Detail}}.
   Critique artefact. Delve into detail by addressing the 30 questions (in six perspectives). The appraiser considers the visualisation from six perspectives using the 30 questions.
   
    \item[\texorpdfstring{\phaseTHREEitem}{3}] {\textbf{Review}}. Lastly, the appraiser reflects on both the holistic critique and the detailed analysis to determine the next steps. The appraiser considers the whole critique, generates an overall score, and decides how to proceed.
\end{itemize}

\smallskip
\noindent To give an example of how the CDS could be used in education, we explain two \textbf{example education tasks that use the CDS}.  An example task could be to ``Design and implement a visualisation poster showcasing a chosen open-source dataset'', another could be to ``design and implement a new interactive visualisation tool of a chosen open-source dataset''. The task is divided into two parts, requiring students to submit two reports along with code and any relevant images, as follows:
\begin{itemize}
    \item Technical Design Plan: This document should include design sketches, consideration of the data story, and the overall layout of the solution. It should outline how the visualisation will effectively communicate the data, explore alternative design approaches, and critically assess their suitability. The Five Design-Sheets (FdS)~\cite{RobertsHeadleandRitsos15_TVCG} method will be used to structure the design process, while the Critical Design Strategy (CDS) will guide reflection and critique.
    \item Visualisation Report: This report will present the final visualisation, accompanied by an in-depth discussion and critique. The CDS framework will be used to structure the evaluation, ensuring a thorough and reflective assessment of the work.
\end{itemize}

\begin{figure*}[h]
\begin{tcolorbox}[title,
    left=0mm,leftupper=0mm,leftlower=0mm,
    top=0mm,bottom=0mm,boxsep=1mm,
    middle=0mm,right=0mm,halign lower=left,
    sidebyside gap=5mm, lower separated=true,
    lefthand ratio=0.2,
     colback=\vignetteColour,colbacktitle=yellow!10!black,
     title={Stage~\numcircledtikz{2} -- Detail} (comprehensive critical evaluation),
     center title,
     before upper*=\fontsize{9}{9.8}\selectfont,
     before lower*=\fontsize{9}{10}\selectfont,
     after lower*=,
   ]
\fontsize{7}{8.2}\selectfont
    \begin{tabular}{l@{\hspace{2mm}}l@{\hspace{2mm}}lc|lll}
    \textbf{Perspectives}&&\textbf{Question on survey}&-2 -1 0 1 2&\multicolumn{3}{l}{\textbf{Range of answers, from poor to good}}\\
    \hline&&&&\\[-.4em]
\textbf{User}&~~\#1&Is suitable for the user and task&\myLikert&Unsuitable&$\leftrightarrow$&Suitable\\
\multirow{4}{*}{\includegraphics[width=\myunitTwo]{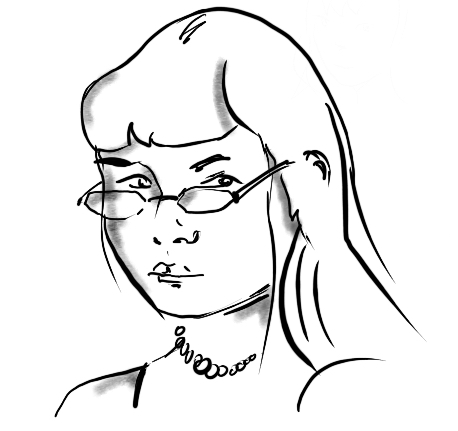}}&~~\#2&Is understandable for user and task to hand&\myLikert&Incomprehensible&$\leftrightarrow$&Understandable\\
&~~\#3&It doesn't require guesswork&\myLikert&Requires guesswork&$\leftrightarrow$&Clear assumptions\\
&~~\#4&Is trustworthy&\myLikert&Distrustful&$\leftrightarrow$&Trustful\\
&~~\#5&Would be useful&\myLikert&Useless&$\leftrightarrow$&Useful\\
   & & & \\[-0.4em]
\textbf{Environment}&~~\#6&It would fit in with other technologies&\myLikert&Wrong setting&$\leftrightarrow$&Right setting\\
\multirow{4}{*}{\includegraphics[width=\myunitTwo]{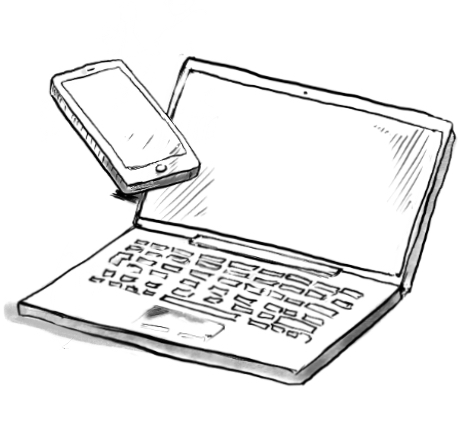}}&~~\#7&Uses suitable technology&\myLikert&Unsuitable technology&$\leftrightarrow$&Right technology\\
&~~\#8&Has appropriate interaction&\myLikert&Unsuitable interaction&$\leftrightarrow$&Appropriate interaction\\
&~~\#9&Its sizing is correct &\myLikert&Unsuitable size&$\leftrightarrow$&Suitable physical size\\
&\#10&Gives a positive ambience&\myLikert&Poor vibe/ambience&$\leftrightarrow$&Positive ambience\\
   & & & & \\[-0.4em]
\textbf{Interface}&\#11&Suitable user interface&\myLikert&Unsuitable GUI&$\leftrightarrow$&Suitable GUI\\
\multirow{4}{*}{\includegraphics[width=\myunitTwo]{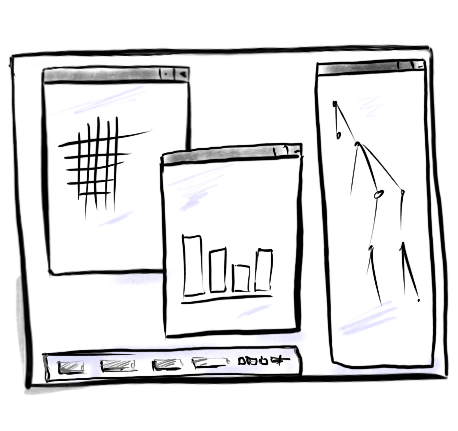}}&\#12&Ergonomic interface&\myLikert&Uncomfortable&$\leftrightarrow$&Ergonomic\\
&\#13&Facets are sized suitably&\myLikert&Poorly proportioned&$\leftrightarrow$&Suitable sized facets\\
&\#14&Interface suitably spaced &\myLikert&Poor facet spacing&$\leftrightarrow$&Relevant spacing\\
&\#15&Suitable quantity of interface parts&\myLikert&Unsuitable facet quantity&$\leftrightarrow$&Suitable facet quantity\\
   & & & & \\[-0.4em]
\textbf{Components}&\#16&Has all necessary components&\myLikert&Missing components&$\leftrightarrow$&All necessary components\\
\multirow{4}{*}{\includegraphics[width=\myunitTwo]{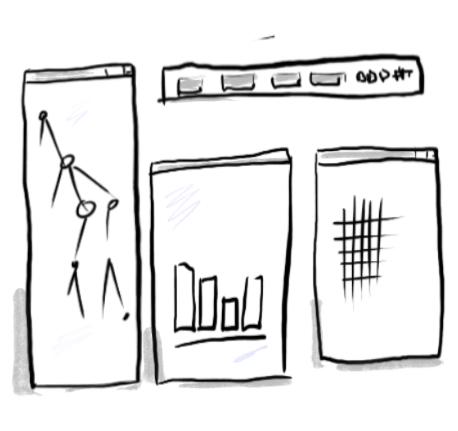}}&\#17&Has all suitable output/view types&\myLikert&Unsuitable types&$\leftrightarrow$&Suitable view types\\
&\#18&Clear relationships between parts&\myLikert&Unclear correspondences&$\leftrightarrow$&Clear view relationships\\
&\#19&Task can be easily performed&\myLikert&Task unfulfilled&$\leftrightarrow$&Task easily performed\\
&\#20&Suitable organisation of components&\myLikert&Poor component layout&$\leftrightarrow$&Good component layout\\
   & & & & \\[-0.4em]
\textbf{Design}&\#21&Inspiring design&\myLikert&Uninspiring&$\leftrightarrow$&Inspiring\\
\multirow{4}{*}{\includegraphics[width=\myunitTwo]{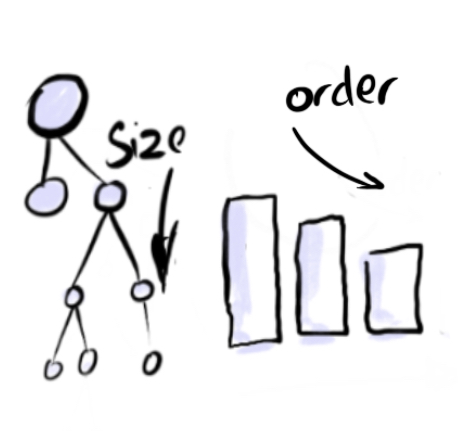}}&\#22&Aesthetic and visually attractive&\myLikert&Unattractive&$\leftrightarrow$&Visually attractive (aesthetic)\\
&\#23&Good composition and space utilisation&\myLikert&Poor layout&$\leftrightarrow$&Good composition\\
&\#24&Suitable coverage of data/underpinning facets/concepts&\myLikert&Unsuitable coverage&$\leftrightarrow$&Suitable coverage\\
&\#25&Clear instructions, labels, legends to give context&\myLikert&Poor labels/legends&$\leftrightarrow$&Suitable legends/labels\\
   & & & & \\[-0.4em]
\textbf{Visual marks}&\#26&Right choice of channels to communicate things clearly&\myLikert&Poor choice of channels&$\leftrightarrow$&Good channel choices\\
\multirow{4}{*}{\includegraphics[width=\myunitTwo]{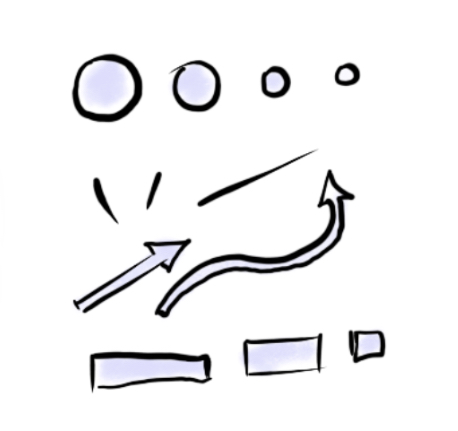}}&\#27&Communicates appropriate relationships/morphisms&\myLikert&Inappropriate mappings&$\leftrightarrow$&Appropriate mappings\\
&\#28&The types of marks used, communicate things well&\myLikert&Inappropriate mark types&$\leftrightarrow$&Suitable mark types\\
&\#29&Components are shown at the right level of abstraction/detail&\myLikert&Poor scale/zoom&$\leftrightarrow$&Good scale/zoom\\
&\#30&Nothing is hidden that shouldn't be hidden&\myLikert&Overplotting&$\leftrightarrow$&Clear display, easy read\\
    \end{tabular} 
\end{tcolorbox}
    \centering
        \vspace{-2mm}\caption{Conduct a comprehensive critical evaluation of the artefact/design. Follow the questions (in the six perspectives: User, Environment, interface, components, design, marks), recording the answers in the Likert scale. Make notes that justify your decisions.  
        }
        \vspace{-2mm}
    \label{Appendixfig:stage2-detail}
\end{figure*}

\textbf{How do use this document.} The detailed questions are designed to assist appraisers in evaluating the design thoroughly. The wording is also intended to guide the teacher in providing a suitable understanding of each section. If the appraiser is confident in understanding the meanings, they can proceed directly to assess each of the three parts and consider the design heuristic questions for each vignette (refer to \cref{Appendixtab:20-first-impression-words-final}, \cref{Appendixfig:stage2-detail} and \cref{Fig:Stage3}).

\definecolor{beige}{rgb}{0.96, 0.96, 0.86}
\definecolor{antiquewhite}{rgb}{0.98, 0.92, 0.84}
\definecolor{oldlace}{rgb}{0.99, 0.96, 0.9}


\section{Stage~\texorpdfstring{\phaseONEitem}{1} -- Overview}

The primary objective of the first stage is to ensure a thorough understanding of the topic and to make holistic assessments of the artefact. Critical thinking necessitates individuals to be ``\textit{well-informed}''~\cite{Facione1990critical}. Individuals should adequately \textbf{prepare} and ensure a thorough understanding of both the challenge and associated data. Data visualisation cannot be pursued without access to data. It crucial to consider the composition of the data and its organisational aspects, such as sparsity and structure. This involves identifying variables, understanding their nature (categorical, ordinal, quantitative, etc.), and recognising the purpose for which the data was collected. Additionally, comprehending the main objective of the visualisation and the intended user tasks is essential. Contextual information, including the creator's intent and the environment in which the visualisation will be utilised, should be understood to ensure ereffective use.

\begin{figure}[h]            
  \begin{tcolorbox}[title,
    left=0mm,leftupper=0mm,leftlower=0mm,
    top=0mm,bottom=0mm,boxsep=1mm,
    middle=0mm,right=0mm,halign lower=left,
    sidebyside gap=5mm, lower separated=true,
    lefthand ratio=0.2,
     colback=\vignetteColour,colbacktitle=yellow!10!black,
     title={Stage~\numcircledtikz{1} -- Overview},
     center title,
     before upper*=\fontsize{8}{9.8}\selectfont,
     before lower*=\fontsize{8}{9}\selectfont,
     after lower*=,
   ]
      
\textbf{Assign a name to the design}:
   \textbf{Summarise essence:}
   \tcblower
     \textbf{Circle 5 (first impression) words:}\vspace{2mm}\newline
     \raggedright clear confusing sensible indifferent clever reliable pointless indistinctive complex organised moderate spectacular useless average bad fulfilling useful fair vague beautiful 
     \end{tcolorbox}
\centering
        \caption{Following adequate preparation, assign name, summarise essence, conduct a holistic critique by selecting five descriptive words.}
        \label{Appendixtab:20-first-impression-words-final}
\end{figure}

To confirm understanding, individuals should \textbf{name} the artefact/design, and summarise its \textbf{essence}. 
The act of naming the design commences the critical thinking process. Crafting a brief, concise title (of two or three words) compels consideration of what is crucial. Similar ideas exist in other design strategies, e.g., with the Five Design-Sheet method~\cite{RobertsHeadleandRitsos15_TVCG} designers are encouraged to name the design categories (on sheet~1) and name their designs (on sheets~2,3 and 4) for the same reason. Other meta-information, such as author name and data, can be added for future reference.
The holistic critique continues, by \textbf{circling five} of the twenty words (\cref{Appendixtab:20-first-impression-words-final}). This task records an preliminary, intuitive assessment. While such instincts can be wrong, they will be reflected upon during  stage~\texorpdfstring{\phaseTHREEitem}{3}.

\definecolor{lightgray}{gray}{0.9}

\section{Stage~\texorpdfstring{\phaseTWOitem}{2} -- Detail}
During the second stage, the aim is to conduct a comprehensive critique (\cref{Appendixfig:stage2-detail}), by considering 30 questions in six perspectives (\textbf{User, Environment, Interface, Components, Design, Visual Marks}), which is recorded using the Likert scale. Documenting the process and decisions can aid in justifying choices. Structure the critique in the order of the stages. 
The six perspectives encourage a top-down approach to the critique, whereby broad aspects are critiqued, like the user and environment, before diving into the specifics of visual elements. The questions are crafted to encourage deep reflection, while the six perspectives help maintain focus on specific design viewpoints. Once a fair evaluation is reached, the overall score can be calculated.

\myTPerspective{4}{-2mm}{pictures/pic1-}{User.\newline}{Critique the artefact or design for user suitability. Empathise with end-user's skills and experience.\vspace{2mm}}

%

When considering these questions, adopt a holistic perspective of the entire tool, visualisation, or system. Put yourself in the shoes of the end-user by empathising with their skills and experiences. Approach the questions from their point of view, keeping their needs and abilities in mind. The appraiser should reflect on the following questions.

\QN{1}{Is suitable for the user and task}{Unsuitable}{Suitable}
Is it suited to the situation and perfectly adapted for its intended purpose~\cite{shneiderman1996eyes}? 
Consider the context in which the design will be used. Does it address the needs and requirements of the target audience effectively? Evaluate whether the design functions as intended and meets the goals set for it. Think about factors such as usability, accessibility, and relevance. Does the design solve the problem it was created for, or does it require adjustments to better align with the needs and expectations of the users? Reflect on whether the design successfully supports the objectives and whether any changes or improvements would make it more effective in fulfilling its purpose.

\QN{2}{Is understandable}{Incomprehensible}{Understandable}  Is the content easily grasped by the end-user? Is it presented in a clear and understandable way?  Consider whether the content of the visualisation is easily understood by the end-user. Does it present the information in a manner that is intuitive and straightforward? Reflect on whether the design makes it easy for the user to grasp the key messages, data relationships, and insights. Is the language, structure, and visual design clear, or could it lead to confusion? Think about the user's potential knowledge level and cognitive load. Does the design cater to different levels of expertise? Is the information logically organised, with a clear flow that guides the user through the content seamlessly? Ensure that the visualisation avoids unnecessary complexity and delivers the message in an easily digestible format.

\QN{3}{Does not require guesswork}{Requires guesswork}{Clear assumptions}
Does it make unwarranted assumptions, possibly relying on domain knowledge, and is this suitable in the context of the user/task? E.g., if intended for public display, clarity and absence of assumptions are required. Does the visualisation rely on assumptions that could confuse or mislead the user? Consider whether it asks the user to make inferences or apply external domain knowledge that might not be reasonable or accessible in the given context. If the visualisation is intended for a general audience or public display, it is crucial that the content is self-explanatory, with no prior assumptions about the user's background or expertise. Reflect on whether the design makes explicit what is being shown, ensuring that all information is clearly presented and no guesswork is necessary. If domain knowledge is required, is it clearly communicated or referenced, so that the user can easily understand the visualisation without making assumptions? Ensure the design is inclusive, accessible, and avoids leaving the user with questions that aren't answered by the visualisation itself.

\QN{4}{Is trustworthy}{Distrustful}{Trustful}
Does the visualisation inspire trust in its data presentation? (Cf. ``Good data visualisation is trustworthy''~\cite{KirkBook2019}). Consider whether the data is presented transparently and reliably. Is it credible and dependable, as described by Meyer et al. \cite{meyer2019criteria}? Reflect on whether the results can be trusted and whether the design instils confidence in the viewer. Is the presentation honest and accurate? Would you feel comfortable recommending this visualisation to others or endorsing it for use? Evaluate whether the visualisation evokes a sense of trustworthiness, ensuring that the information it presents can be confidently relied upon.

\QN{5}{Would be useful}{Useless}{Useful}
Does the visualisation serve a practical purpose in its context? Consider whether it provides value to the user or the task at hand. Is it likely to be used effectively, and does it enhance the understanding or decision-making process? Reflect on the context in which the visualisation will be applied. If for a specific project, presentation, or public display, and assess if it is appropriate and beneficial for that setting. Does it help solve a problem, answer a question, or provide insights that would be difficult to obtain otherwise? Is it intuitive enough to engage the user and facilitate interaction or interpretation? In short, does it add value to the task or situation, or does it fall short in meeting its intended goals?

\myTPerspective{4}{-2.1mm}{pictures/pic2-}{Environment.}{Assess its suitability for proposed environment. Critique overall scenario, setting, and technology. Conversely, environmental obstacles could impact the artefact.} 

Evaluate if the artefact is appropriate for the intended environment. Assess the entirety of the scenario: setting, technology, and platform, whatever the environment such as print form, e-magazine, smartphone, tablet, desktop or powerwall.
With the environment perspective, you need to imagine the environment, technology that will be utilised, circumstances of its use and how a user would operate it or interact with it in that situation. It is all about appropriateness for the environment.

\QN{6}{It would fit in with other technologies}{Wrong setting}{Right setting}
Does the visualisation align with and complement the other technologies and systems in its intended environment? Consider if it integrates smoothly and facilitates effective interoperability. For example, a static display might be ideal for an e-book, while a 3D model would be better suited for an immersive head-mounted display. The design may be intended for a printed magazine or an interactive desktop tool; both could be appropriate in their respective contexts. However, an interactive tool may not work well in a print magazine, though a QR code that launches a 3D view on a head-mounted display could be a suitable alternative. Additionally, a 3D stereo design might be less appropriate for print media. Reflect on whether the visualisation is appropriately adapted to the setting and if it enhances the user experience within that specific context.

\QN{7}{Uses suitable technology}{Unsuitable technology}{Right technology}
Is the technology used in the design appropriate and well-matched to the intended purpose, environment, and user needs? Does it consider ergonomic factors? For instance, a static display may be ideal for an e-book, while a 3D environment would be better suited for an immersive head-mounted display. Does the chosen technology enable the intended actions effectively? For example, can the user perform tasks such as ordering, organising, or scaling using the provided technology? Consider whether a large-scale visualisation would work well on a small mobile screen or if it would lose its effectiveness. Reflect on whether the technology is the right fit for the context and task at hand.

\QN{8}{Has appropriate interaction}{Unsuitable interaction}{Appropriate interaction}
Can you perform the desired actions within the environment using the provided technology, and how well does it integrate with other technologies in the setting? The term interaction should be interpreted based on the specific context. Rather than only evaluating if it's a Human-Computer Interaction (HCI) tool (e.g., using a mouse), assess if the level and type of interaction suit the environment. For example, interacting with a physical book (turning pages or adjusting its position) is appropriate without needing a computer. In a different scenario, physical interaction with a powerwall might involve moving closer or farther away, but environmental factors could influence this. Similarly, using a dashboard with a mouse, pen interface, or voice commands should be evaluated in terms of its suitability for the given context. Does the interaction functionality meet the requirements of the environment? Is it organised effectively? Can you easily perform actions like scaling or zooming when necessary? Even in cases where there is no direct human/computer interaction, such as with a poster where people move closer or farther away, the interaction may still be appropriate.

\QN{9}{Its sizing is correct}{Unsuitable size}{Suitable physical size}
Is the size of the output appropriate for its intended use and context? Consider whether the size allows the user to view all the necessary information clearly and comfortably. If the size is too small, critical data may be obscured or difficult to interpret. Conversely, if the size is excessively large, it may overwhelm the user, causing difficulty in viewing all the content at once or forcing unnecessary scrolling or zooming. The size should support effective data presentation while remaining user-friendly and accessible. Is it properly scaled to the device or medium in which it's displayed, whether that be a small mobile screen, a large display, or printed material?

\QN{10}{Gives a positive ambience}{Poor vibe/ambience}{Positive ambience}
Does the artefact convey the intended atmosphere or feeling to the user? Does it create an engaging, pleasant, and welcoming experience, enhancing the interaction rather than hindering it? Consider how the design's visual appeal, layout, colours, and interactive elements contribute to the overall user experience. Does it align with the context in which it's being used? For example, a financial dashboard may benefit from a clean, professional look, while an educational visualisation could use more engaging and playful aesthetics. Does the design help create a positive, motivating, or inspiring environment for the user, supporting the overall goal of the tool or presentation?

\myTPerspective{4}{-2.1mm}{pictures/pic3-}{Interface layout.\newline}{Consider the organisation of the interface and the graphical user interface (if applicable) assessing its suitability for the intended purpose.}

Consider the overall interface layout and how the user interacts with the visuals. For example, a computer screen might display visuals, with the user interacting through a keyboard and mouse, potentially accompanied by sound. On a tablet, users may swipe with their fingers, while in a virtual reality environment, physical movement is the primary form of interaction. Alternatively, a print magazine requires the user to physically engage with the content, such as flipping through pages or even cutting out articles with scissors.

\QN{11}{Suitable user interface}{Unsuitable GUI}{Suitable GUI}
Is the user interface appropriate for the task at hand? Consider the specific features it offers, such as menus, buttons, drag-and-drop functionality, and visual programming interfaces. Should the interface be static or dynamic? Does it allow for transparency, dynamic queries, direct manipulation, or data querying? If the interface uses tabbed, cascaded, or tiled layouts, are these suitable for the context?
Evaluate whether the provided interface meets the task requirements and is fit-for-its-purpose. Does it allow users to query data when needed? Are tabbed, cascaded, or tiled interfaces appropriate for the visualisation? Consider whether the method of interaction (e.g., visual programming of data flow, defining queries by code, or scrolling to locate information) is suitable for the user and task. Does the interface support efficient interaction and data exploration in a way that aligns with the intended use?

\QN{12}{Ergonomic interface}{Uncomfortable}{Ergonomic}
Is the interface designed in an ergonomically-friendly way, ensuring ease of use and comfort for the user? Consider whether users have the necessary interface components easily accessible and within reach. For a hand-held physical device, does it fit comfortably in the user's hand? Is it easy to hold and use for extended periods? Evaluate whether the interface is well-structured, unobtrusive, and user-friendly. For example, a drag-and-drop command may create an ergonomic interface by making interaction intuitive and quick, whereas an interface requiring multiple menus or numerous button clicks may be less comfortable and more cumbersome to navigate. Similarly, selecting objects in a 3D VR world may initially be fine, but long periods of interaction could lead to fatigue. Does the interface facilitate seamless, comfortable use over time, or does it require adjustments that might affect user comfort and engagement?

\QN{13}{Facets are sized suitably}{Poorly proportioned}{Suitable sized facets}
In visual displays, different facets or components are often used to present information. The size of these facets plays an important role in ensuring the clarity and effectiveness of the display. Are the sizes of these facets appropriate for the task at hand? For instance, when comparing two datasets (e.g., data A vs. data B), the size allocated to each dataset should be relatively equal to avoid unintentionally biasing one over the other. Similarly, in a web viewer where a central display area is accompanied by advertisements, it's crucial to consider if the size of the adverts is appropriate in relation to the main content. If the adverts are too large or too small compared to the central display, it may disrupt the balance and overall user experience. Therefore, sizing must be carefully considered to maintain harmony and facilitate effective communication. Does the size of each facet align with its intended purpose and contribute to a clear, well-organised visual presentation?

\QN{14}{Interface suitably spaced}{Poor facet spacing}{Relevant spacing}
Does the interface make effective use of space within the layout, facets, or frames? Consider how white space is employed, or used appropriately. Spacing can improve clarity, enhance readability, and highlight key elements, contributing to a calm and organised visual aesthetic, as advocated by Dieter Rams~\cite{DeiterRams} -- and the calming nature of simplicity. The Gestalt psychologists also emphasised the importance of spacing (through the law of proximity). Their work describes that items placed close together are perceived as connected and related, while elements that are spaced further apart appear less connected and more distinct from one another~\cite{ware2012information}. However, it is important to avoid excessive space, as this may create unnecessary gaps that disrupt the visual flow and hinder user comprehension. For example, too much space between related elements can confuse the viewer and undermine the intended relationships between them, violating the Gestalt principle of proximity. Effective spacing should maintain a balance, providing enough room to differentiate between elements while ensuring that related items remain visually connected. Does the spacing between facets and content help organise information efficiently, allowing the user to navigate and interpret the design with ease?

\QN{15}{Suitable quantity of interface parts}{Unsuitable facet quantity }{Suitable facet quantity} 
Is the number of interface components or facets appropriate for the task at hand? An excessive number of windows or facets can overwhelm the user, adding unnecessary complexity and making it difficult to focus on the key elements. On the other hand, too few facets may result in the lack of critical information, limiting the effectiveness of the display. It is essential to strike a balance where the quantity of facets supports clarity and functionality without causing cognitive overload. Consider how the interface components work together to provide a coherent user experience, ensuring that every part serves a distinct purpose and contributes to the overall task.

\myTPerspective{4}{-2mm}{pictures/pic4-}{Components.}
{Components are specific visual elements or depictions that can be identified and isolated for individual consideration. Identifying and understanding these components is essential for analysing the structure and effectiveness of the design/ visualisation as a whole.}

Each component within a design, artefact, or visualisation has a specific role and contributes to the overall presentation, conveying particular information to the viewer. These components can take various forms, such as charts, graphs, tables, icons, or other visual data representations~\cite{ChenETAL_2021}. They also include elements like menus, labels, and help information. Components may be arranged in different formats, such as a grid layout (a matrix of small, multiple views) or within a tabbed interface~\cite{al2019towards}. These components serve as the building blocks of the design, often placed within frames or windows. Examples include bar charts, timelines, treemaps, scatterplots, and others. Each component is distinct and identifiable, contributing to the overall function and user experience of the visualisation.

\QN{16}{Has all necessary components}{Missing components}{All necessary components}
Does the design include all the essential components needed to effectively convey the information? Are any crucial elements missing or unavailable? For example, if the design is intended to display temporal databut a timeline is absent, it may hinder the user's ability to interpret the data. Similarly, certain options, such as the `delete' function, may be grayed out or unresponsive, preventing users from performing necessary actions. It's also possible that a component might be obstructed or hidden behind another element in the interface, leading to its absence from the user's view or interaction. The completeness of the interface is essential for ensuring a seamless and intuitive experience, where all required functionality is accessible and visible to the user.

\QN{17}{Has all suitable output/view types}{Unsuitable types}{Suitable view types}
Does the design incorporate the appropriate output or view types to support the task and enhance interpretation? The choice of visualisation type is crucial for facilitating the use's understanding and task execution. For instance, continuous data is best represented using a line graph, while categorical data is typically shown using a bar chart. The visualisation type should align with the goal of the task; whether it's comparison, identification, or interpretation of values. For example, if the goal is to compare two items, placing them close together in the design can aid comparison. However, if items are too far apart, comparison becomes difficult, making the visualisation type unsuitable. Similarly, highly aggregated visualisations or those with large data bins may make it difficult to extract exact values. Therefore, it's important to assess whether the selected visualisation type, such as a bar chart, line graph, or image, is best suited for the intended task and the data being presented.

\QN{18}{Clear relationships between parts}{Unclear correspondences }{Clear view relationships}
Are the relationships between different components of the display clearly communicated?  Is the legend or key clearly associated with the visual elements it describes? Does the title logically correspond to the content? In any visualisation, parts typically relate to one another, either implicitly or explicitly, so does the design make these connections clear? For example, in a multiple-view visualisation, is the relationship between different views immediately obvious? These linkages could be made apparent through annotations, or colouring, or visual links (so called meta-visualisation techniques~\cite{weaver2005aSitu}).
Relationships can be conveyed through various design techniques, such as proximity (in line with Gestalt principles~\cite{Munzner2014Book,Ware12}), bounding areas, explicit arrows, or other visual cues like speech bubbles to reinforce connections between elements.

\QN{19}{Task can be easily performed}{Task unfulfilled}{Task easily performed}
Can you perform the task you need to do effectively? Consider the specific task and how the user interface and its components facilitate completing it (see Shneiderman~\cite{shneiderman1996eyes}). For instance, if the task involves telling a story of change over time, does the visualisation use a plot that clearly shows the progression over time? If the task requires illustrating three distinct phases, are these phases represented clearly and visually distinct within the plot? Are the components, such as plots or charts, fulfilling their intended purpose in a way that supports the task at hand?
For example, if you're using a dashboard, is it clear what actions the dashboard is designed to perform? Can you easily take the necessary actions within the interface? If a pie chart is chosen, does it effectively convey the data and align with the task's goals~\cite{Kosara2019Pie}? Each component serves a unique purpose—one component may illustrate time, while others may detail data within specific phases. Are these visual components intuitively presented, ensuring users can easily interpret the information without confusion?
Additionally, does the interface design adhere to usability principles, such as providing the right level of detail without overwhelming the user? The effectiveness of the task hinges not only on the appropriateness of the chosen visualisation types but also on how well the user can interact with the interface components. This includes ensuring that the controls (mouse movement, button presses, touch gestures, or 3D interface elements) are intuitive and enable users to complete the task smoothly (see also heuristics \#11 to \#15).

\QN{20}{Suitable organisation of components}{Poor component layout}{Good component layout}
Is the arrangement and ordering of the components in the visualisation effective and logical? For instance, if there's a timeline alongside other visual elements, where should it be positioned for optimal clarity (e.g., at the top or bottom of the display)? When comparing two items, what is the best positioning? Should they be placed next to each other, stacked vertically, or positioned side by side for easy comparison~\cite{GleicherComparisonETAL2011}? For example, placing two bar charts close to one another allows for a more straightforward comparison, in line with the Gestalt principle of proximity~\cite{Munzner2014Book,Ware12}. If components are spaced too far apart, it can make comparison more difficult, potentially hindering the user's ability to extract meaningful insights. Therefore, is the component layout designed in a way that enhances comparison and clarity? Does the structure align with the task and the user's ability to interpret the information efficiently?

\myTPerspective{5}{-2mm}{pictures/pic5-}{Design.\newline}{
Design encompasses organising any part of the system, which involves considerations like colour balance, item alignment, and styling.}

Good design encompasses the effective organisation of all elements within a system. For traditional visualisation on a desktop computer, this involves considering aspects such as the balance and use of colours, the alignment of components, styling, and other visual details that contribute to a cohesive and intuitive experience.

\QN{21}{Inspiring design}{Uninspiring}{Inspiring}
Did the design immediately capture your attention, leaving you thinking, ``Wow''? Does it motivate you to apply its concepts to future projects? Does it align with established design principles, such as the Golden Rules~\cite{ShneidermanETALBookDesign} or Dieter Rams' good design principles~\cite{DeiterRams}? Upon encountering the design, do you feel an immediate urge to interact with it and explore its features?
Good design is made up of many elements, and in this case, we focus specifically on the layout. Are design elements organised consistently and thoughtfully? Are comparable items placed close together, making comparisons intuitive? Are the colours used effectively, without overwhelming the user? Does the design make you want to share it on social media because of its appeal?

\QN{22}{Aesthetic and Visually Attractive}{Unattractive}{Visually Attractive (Aesthetic)}
While the perception of attractiveness can be subjective, research shows that people tend to favour designs that are balanced, harmonious, and well-proportioned~\cite{Ware12}. A visually appealing design should feel cohesive and comfortable to the eye, avoiding visual clutter or overwhelming elements. Does the design utilise a colour palette that is appropriate for the topic and context, without using an excessive number of colours that might distract or confuse the viewer~\cite{Harrower2013colorbrewer}? Is the colour scheme thoughtfully selected, enhancing the content rather than competing with it~\cite{BorlandTaylor2007}? Is the colour combination web-safe, or accessible~\cite{Harrower2013colorbrewer}?
In addition, consider whether the visual appeal of the design would make you confident in presenting or using it in a professional setting. Would you feel comfortable sharing it with clients, colleagues, or stakeholders? An aesthetically pleasing design can foster trust and engage users more effectively, encouraging them to interact with the content and absorb the information presented. Does the overall look of the visualisation inspire positive reactions or convey professionalism and attention to detail?

\QN{23}{Good composition and space utilisation}{Poor layout}{Good composition}
Does the design demonstrate a thoughtfully arranged composition, with well-organised elements, colours, and visual components that clearly communicate the intended message or information? Consider how the individual components are arranged: Are they placed strategically to create a harmonious and efficient layout? Is space optimally utilised to avoid clutter or waste? For example, in a book, generous gutter spacing ensures a comfortable reading experience. A magazine layout might feature two main columns with picture insets flowing between gutters, creating a dynamic and visually balanced structure. Some publishers allow text to flow around images, while others prefer fully justified text for a clean, uniform look. Does it fulfil the requirements of the output? E.g., a poster display may be required to be a certain size? Does it allow the quality of output at that resolution? E.g., if a pixel format is used, then it may not scale well.
In some cases, overlapping or cascading elements might be appropriate for certain designs, but excessive overlap can lead to a chaotic and disorganised appearance, which may detract from the overall clarity of the presentation~\cite{ShneidermanETALBookDesign} (see also \#30). A well-composed layout ensures that every element serves a purpose while maintaining visual clarity and balance, guiding the user’s attention effectively through the content.

\QN{24}{Suitable coverage of data/underpinning facets/concepts}{Unsuitable coverage}{Suitable coverage}
Does the design effectively display all the necessary data? Is the quantity of data presented appropriate for the task? Is the data aggregation method used correct and appropriate for the context? For example, is it suitable to compress an axis in some cases, or could this approach potentially confuse the user? Does the visualisation properly represent all required transformations and relationships in the data?
Is sparse or missing data clearly represented in a way that doesn't mislead the viewer? Does the visualisation convey the full story it aims to tell, presenting a comprehensive picture? Moreover, is the chosen approach for data representation the most effective for the intended narrative or message, ensuring clarity and understanding?

\QN{25}{Clear instructions, labels, legends to give context}{Poor labels/legends}{Suitable legends/labels}
Is the contextual information provided sufficient to help users clearly interpret the displayed data?
For example, if a visualisation is labelled ``oil usage'' the meaning could be misunderstood if the context suggests ``cooking oil'' rather than a broader term. Similarly, a vague instruction like ``move forward'' could create confusion, where more specific phrasing such as ``move forward five steps'' or ``move forward 5 meters'' may be necessary depending on the context. If no labels or legends are provided, is their absence justifiable? Can they be added to improve clarity? Are the existing labels, legends, titles, or other explanatory elements~\cite{DykesETAL2010} accurate and sufficient in helping to explain the data? If these elements are missing, does it leave users uncertain about what the visualisation is conveying, or does it make sense to omit them? For example, scatterplots often omit labels on individual points. Is this omission acceptable, or does it hinder understanding? Would adding labels improve clarity, and if so, how could they be positioned and formatted to avoid overcrowding and maintain readability? The number, placement, and clarity of labels are crucial for effective communication; are these considerations addressed in your display?

\myTPerspective{4}{-2mm}{pictures/pic6-}{Visual marks.}{Visual marks encompass graphical elements like lines, shapes, colours, and textures~\cite{Bertin1983}. Their layout should avoid overcrowding and ensure accurate representation of data. Correct data morphisms are essential for conveying information effectively.}

In this context, the focus is on the appropriate use of graphical marks to effectively represent data. Are the correct marks chosen for the data type, and are they positioned accurately with the right attributes to clearly communicate the intended message? Additionally, evaluate whether any design elements hinder clarity or overcomplicate the visualisation.
Graphical marks are fundamental properties of the visual system, often referred to as retinal variables by Bertin~\cite{Bertin1983}. These marks encode data through attributes such as size, orientation, colour, texture, and transparency. They can range from basic elements like lines, polygons, and circles, to more complex pictures, icons, or multidimensional glyphs~\cite{WardTaxonomyGlyphPlacement2002}. It is important to assess how these marks are arranged—whether they are appropriately placed, not overcrowded, and whether they effectively convey relationships between the data. Bertin's concept of retinal variables highlights how certain graphical marks are more suitable for specific tasks depending on the data being represented. For example, size is particularly effective for conveying quantitative values: larger objects can represent larger values, making it easier for viewers to intuitively grasp relative magnitudes. Similarly, certain visual attributes, such as colour or shape, may be better suited for identifying or distinguishing between categories or groups within the data. However, not all retinal variables can be mapped to every type of data or understood visually in the same way. For instance, while colour can be used to represent categorical data, it may not effectively convey this information, and using it for that purpose could lead to confusion. Bertin’s theory emphasises the importance of selecting the right visual attributes that align with the nature of the data and the task at hand, ensuring that the chosen graphical marks effectively communicate the intended message without introducing misinterpretation or ambiguity. Furthermore, consider if any visual elements are redundant or unnecessary, such as ``chartjunk''\cite{tufte1983visual}, and how the design aligns with established visualisation principles\cite{Ware12}. Keep in mind that not all extra visual elements are detrimental; in some cases, they can enhance the memorability and impact of the visualisation~\cite{borkin2013makes}.

\QN{26}{Right choice of channels to communicate things clearly}{Poor choice of channels}{Good channel choices}
Different channels, such as position along an aligned scale, shape, size, orientation, and colour, are used to encode data values and convey information. How well are these channels being utilised? Do they effectively match the data being presented? These channels engage various sensory modalities: for instance, colour is perceived through sight, while vibration can be felt, as seen in vibrotactile interfaces. Are the chosen channels compatible with the user's sensory experience? Depending on the design context, could audio be a more effective medium? Is it the best choice in this scenario, or might it be overwhelmed in a noisy environment?
When selecting channels, is the environment considered? For example, is the visual information readable in a well-lit or dim setting? If you are using audio, is it clear enough for users to discern the message? Are there alternatives that might be more inclusive or accessible for those with sensory impairments? Does the chosen channel ensure that the information is easily understood by the intended audience, and are these channels contributing to a more inclusive experience~\cite{kim2021accessible}? Ultimately, do the channels enhance the user's comprehension without introducing confusion?

\QN{27}{Communicates appropriate relationships/morphisms}{Inappropriate mappings}{Appropriate mappings}
Is the mapping, or transfer function, aligned with the data, and does it accurately reflect the underlying relationships? Are the choices made for visualising the data appropriate in terms of the relationships they intend to convey? For example, when mapping continuous data, is it represented in a continuous way, or does using discrete categories or binning distort the data's true nature? If categories are represented, is the grouping meaningful, or does it obscure important variations? For instance, in a bar chart, does the grouping of values into ranges (e.g., 1-5, 6-10, 11-15) appropriately reflect the characteristics of the data, or does it introduce unnecessary abstraction?
How do the mappings affect user understanding? Are they intuitive and clear? For instance, does colour or size effectively represent data values, or are these elements misleading? Can the user easily make connections between data points and visual elements?
Does the design ensure the right balance between simplicity and accuracy? Does the visualisation help the user draw meaningful conclusions without overcomplicating the task? Does the mapping enhance the clarity of relationships, trends, or patterns within the data?

\QN{28}{The types of marks used, communicate things well}{Inappropriate mark types}{Suitable mark types}
Are the marks used in the presentation appropriate for the data and the task at hand? In a scatterplot, for instance, different types of marks (such as points, circles, triangles, or lines) are employed to represent data. Are these marks the right choice for effectively conveying the relationships and characteristics of the data? Are symbols, lines, or areas used in a way that makes sense within the context of the visualisation? Do attributes such as line style (dotted, dashed, or solid) or colour (representing categories, values, or gradients) help to communicate the intended message?
Is it clear what each mark represents, and is there any potential confusion caused by using similar or overly complex marks? Do the marks enhance the clarity of the visualisation or do they distract from the main message? For example, are the shapes or colours intuitive to users, and are they consistent across the visualisation? Consider if certain marks are overused or redundant and if alternative marks might communicate the data more clearly. Does the choice of marks reflect the nature of the data, such as using continuous marks for continuous data and categorical marks for discrete data?
Key questions to consider: Are the chosen marks appropriate for the data type? Are any visual elements unclear or difficult to interpret? Do the marks and their attributes support the task effectively?
Is there consistency in the use of marks throughout the visualisation?

\QN{29}{Components are shown at the right level of abstraction/detail}{Poor scale/zoom}{Good scale/zoom}
Are the visual marks presented at the appropriate level of abstraction or detail? For instance, are the sizes, shapes, or colours of the marks used at the correct scale to effectively communicate the data? Would zooming in on specific visual marks or adjusting their size enhance clarity, allowing the user to focus on finer details without losing context? Are there any non-linear zoom options, like distortion views, that could make the data more interpretable while maintaining the overall structure of the visualisation? Should certain visual marks, such as lines or points, be emphasised by scaling them larger or smaller for better comprehension? Consider whether the marks are too abstract or too detailed for the task at hand and whether their level of granularity is consistent and appropriate for the data being communicated.

\QN{30}{Nothing is hidden that shouldn't be hidden}{Overplotting}{Clear display, easy read}
When plotting data, overlapping points can obstruct each other, making interpretation difficult. Is the arrangement of marks appropriate, or are they too close together, resulting in visual clutter? Could a different transfer function improve clarity by either separating the points or aggregating them in a more meaningful way? Consider whether filtering options might help reduce overplotting and provide a clearer view of the data.
In 3D visualisations, occlusion can occur when objects are hidden behind others, or sounds overlap in an audio space, interfering with clarity. Is this type of occlusion acceptable, or could it be managed better? Would a different layout, such as repositioning elements or introducing transparency, help mitigate the issue? Additionally, there may be cases where intentional partial occlusion is used for specific purposes: Does this make sense in the context, or does it hinder understanding? Evaluate whether partially hidden elements improve or hinder the user’s ability to interpret the visualisation effectively.

\begin{figure}[h]            
  \begin{tcolorbox}[title,width=\linewidth,
    left=0mm,
    middle=0mm,right=4mm,
    lower separated=true,
    lefthand ratio=.4,
    lefthand width=.2\linewidth,
     colback=\vignetteColour,colbacktitle=yellow!10!black,
     title={Stage~\numcircledtikz{3} -- Review},
     center title,
     before upper*=\fontsize{8}{9}\selectfont,
     before lower*=\fontsize{8}{9}\selectfont\begin{tabular}{@{}p{.5\linewidth}p{.5\linewidth}@{}},
     after lower*=\end{tabular},
   ]
      \vspace{-3mm}\textbf{Create score.}\hspace{1cm}
      \textbf{Reflect on parts:}\hfill
      \tcblower
       \textbf{Improvements and next steps:}&\\
     \end{tcolorbox}
\centering
        \vspace{-3mm}\caption{The final stage involves synthesising the various perspectives and insights, gathered throughout the critique.}\vspace{-3mm}
        \label{AppendixFig:Stage3}
\end{figure}

\section{Stage~\texorpdfstring{\phaseTHREEitem}{3} -- Review}
The goal of the final stage (\cref{AppendixFig:Stage3}) is to synthesise key findings and observations, turning these insights into actionable steps that contribute to refining and improving the design or artefact. The first step involves calculating a score by summing the Likert scale values and reflecting on each component of the critique, such as the name, essence, and six perspectives. While the average score from the Likert scale can provide useful guidance, it may not fully capture the depth of the critique and could be misinterpreted, as it masks the complexities of individual aspects. Therefore, it's crucial to interpret the score alongside other insights. Review the six perspectives and the 30 questions, identifying any particularly noteworthy strengths or weaknesses. Highlight areas that require improvement and pinpoint elements that should be enhanced.

Once the critique has been thoroughly assessed, the next step is to decide on appropriate actions. A redesign may be necessary to address the identified issues and improve the design. This could involve refining the layout, adjusting visual elements, enhancing usability, or incorporating user feedback. The next steps should include developing a detailed plan for the redesign, such as conducting further research, gathering more user input, creating prototypes, and implementing changes iteratively. Establishing clear objectives for the redesign is essential, and regular evaluations should be conducted to ensure that the new design effectively addresses the identified issues, improving overall usability and user experience.

\balance

\clearpage
\renewcommand{\vgtc@preprinttext}{}  
\thispagestyle{empty}
\includepdf[pages=-,pagecommand={}]{CDS_V4_longVersionVIS2025.pdf}

\clearpage
\renewcommand{\vgtc@preprinttext}{}  
\thispagestyle{empty}
\includepdf[pages=-,pagecommand={}]{CDS_V4_sheetVersionVIS2025.pdf}
\end{document}